\documentclass[nonacm,sigplan,screen]{acmart}

\setcopyright{none}
\renewcommand\footnotetextcopyrightpermission[1]{}

\usepackage{amsmath}
\usepackage{algpseudocode}
\usepackage{graphicx}
\usepackage{subcaption}
\usepackage{multirow}
\usepackage{diagbox}
\usepackage{threeparttable}
\usepackage{booktabs}
\usepackage{makecell}
\usepackage{pifont}
\usepackage{enumitem}
\usepackage{bbding}
\usepackage[ruled,vlined]{algorithm2e}
\SetKwProg{Fn}{Function}{:}{end}

\AtBeginDocument{%
  \providecommand\BibTeX{{%
    \normalfont B\kern-0.5em{\scshape i\kern-0.25em b}\kern-0.8em\TeX}}}

\begin{document}

\newcommand\blfootnote[1]{%
\begingroup
\renewcommand\thefootnote{}\footnote{#1}%
\addtocounter{footnote}{-1}%
\endgroup
}

\title{PDScope: Breaking the Prefill-Decode Seesaw for Memory-Bound MoE Inference
}

\author{Enda Yu}
\affiliation{%
  \institution{National University of Defense Technology}
  \city{Changsha}
  \country{China}
}
\email{yuenda@nudt.edu.cn}

\author{Dezun Dong}
\affiliation{%
  \institution{National University of Defense Technology}
  \city{Changsha}
  \country{China}
}
\email{dong@nudt.edu.cn}

\author{Zhaoning Zhang}
\affiliation{%
  \institution{National University of Defense Technology}
  \city{Changsha}
  \country{China}
}
\email{zhangzhaoning@nudt.edu.cn}

\author{Xiaojian Yang}
\affiliation{%
  \institution{National University of Defense Technology}
  \city{Beijing}
  \country{China}
}
\email{yangxj@nudt.edu.cn}

\author{Weiling Yang}
\affiliation{%
  \institution{National University of Defense Technology}
  \city{Changsha}
  \country{China}
}
\email{w.yang@nudt.edu.cn}

\author{Shaonan Ma}
\affiliation{%
  \institution{Qiyuan Laboratory}
  \city{Beijing}
  \country{China}
}
\email{mashaonan@qiyuanlab.com}

\author{Dongsheng Li}
\affiliation{%
  \institution{National University of Defense Technology}
  \city{Changsha}
  \country{China}
}
\email{dsli@nudt.edu.cn}

\author{Yongwei Wu}
\affiliation{%
  \institution{Qiyuan Laboratory}
  \city{Beijing}
  \country{China}
}
\email{wuyw@tsinghua.edu.cn}

\author{Xiangke Liao}
\affiliation{%
  \institution{National University of Defense Technology}
  \city{Changsha}
  \country{China}
}
\email{xkliao@nudt.edu.cn}


\begin{abstract}


Local deployment of MoE-based LLMs on CPU-GPU nodes is attractive for cost and privacy reasons, yet limited GPU memory forces most experts into CPU memory. Prefill is then compute-bound and benefits from aggressive GPU loading, while decode is I/O-bound and requires conservative PCIe transfers. These demands cannot be amortized under low batching---typical of local assistants and document tools---forcing phase-agnostic schedulers into a seesaw: favoring prefill's aggressive loading injects unmaskable I/O that stalls decode, while favoring decode's conservative transfers starves prefill's GPU of work.
We observe that expert routing is predictable in phase-specific ways, allowing us to decouple the two phases and optimize prediction, scheduling, and execution independently for each phase. PDScope predicts expert activations to guide placement and prefetching, adapts these policies to each phase's bottleneck via cross-layer cost modeling, and compiles decisions into overhead-free pipelines using a static computation graph, cache prefetching, and zero-copy asynchronous I/O.
On mainstream MoE models, PDScope achieves 2.5–10.3× higher prefill throughput and 2.1–6.1× higher decode throughput than existing systems, showing that phase-aware co-design unlocks efficient MoE inference under memory constraints and extends naturally to multi-batch settings.

\end{abstract}

\maketitle

\begin{figure}[t]
  \centering
  \begin{minipage}{\linewidth}
    \centering
    \includegraphics[width=\linewidth]{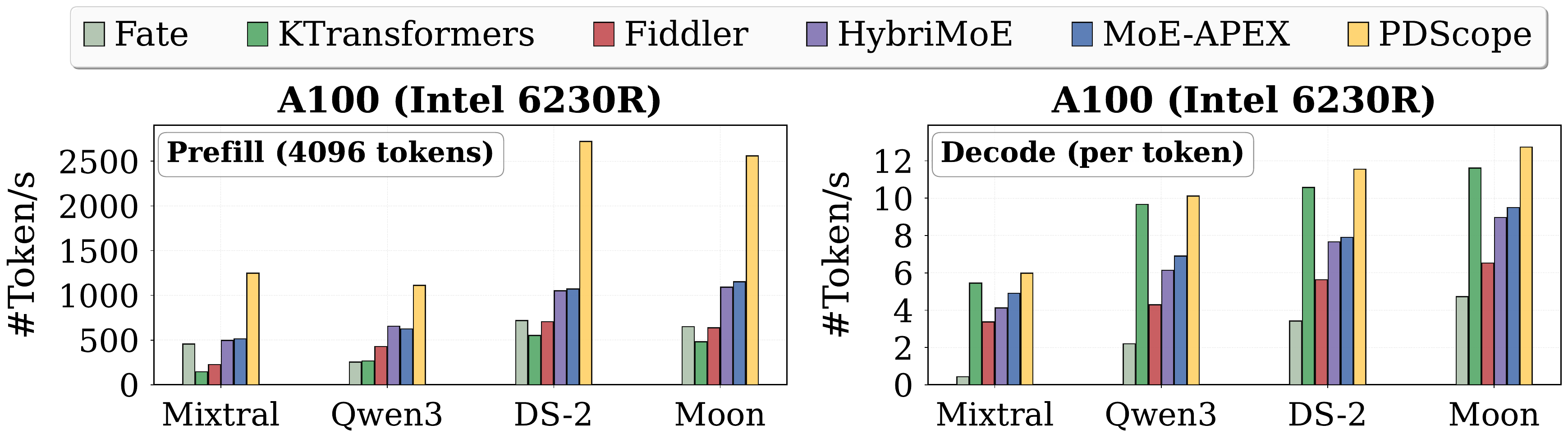}
    \captionof{figure}{Throughput comparison of prefill (4096 tokens) and decode (per token) on the A100 platform (\S\ref{expset}).}
    \label{figure1}
  \end{minipage}
 
  \begin{minipage}{\linewidth}
    \centering
    \includegraphics[width=\linewidth]{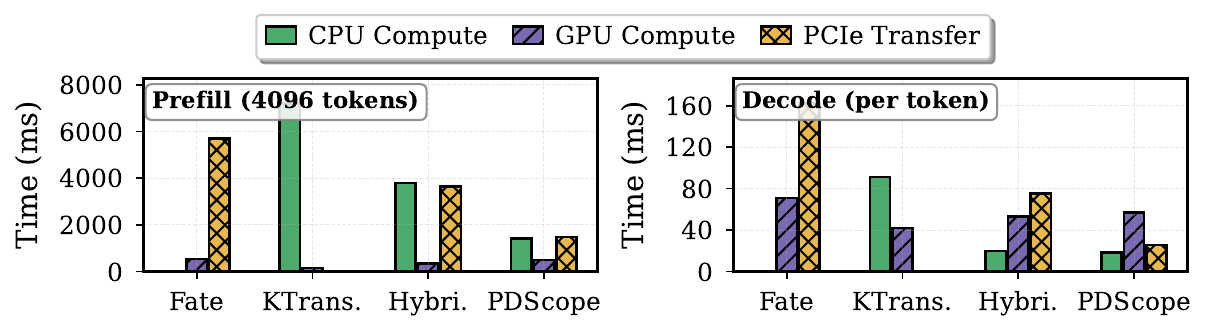}
    \captionof{figure}{   Breakdown of inference overhead for DeepSeek-V2-Lite on the A100 platform. Baselines exhibit a seesaw: Fate-style on-demand loading boosts prefill GPU compute but spikes decode PCIe overhead; kTransformers-style static placement tames decode I/O at the cost of idle prefill GPU. PDScope breaks this trade-off by balancing both metrics.}

    \label{figure1b}
  \end{minipage}
\end{figure}

\section{Introduction}
\label{sec:introduction}

The sparse activation of MoE models~\cite{kimi3,ds4,llama4} makes them ideal for local deployment on CPU-GPU servers in cost-sensitive or privacy-prioritizing scenarios. However, limited GPU memory forces most expert parameters to reside in CPU memory, causing the two inference phases to exhibit drastically different bottleneck characteristics.

The prefill phase processes input tokens in parallel, with performance determined by the fraction of tokens processed on the GPU—higher GPU computation translates to greater throughput. The decode phase generates tokens autoregressively, with performance dominated by PCIe transfer latency \cite{diff,InstAttention,Zero-infinity,PowerInfer}—lower I/O overhead means reduced latency. 
These two phases place conflicting demands on system resources: prefill benefits from aggressive expert loading to maximize GPU utilization, while decode requires conservative I/O strategies to avoid transfer bottlenecks. This trade-off creates an inherent seesaw effect: optimizing for one phase inevitably degrades the other, as any static resource allocation cannot satisfy both phases simultaneously.

Existing systems explore different trade-off points with unified scheduling logic. Static placement strategies~\cite{ktransformers,llamacpp,sglang,vllm} (e.g., kTransformers~\cite{ktransformers}) keep hot experts (those processing more tokens) resident on the GPU, prioritizing decode stability but leaving GPU compute underutilized during prefill. On-demand loading strategies~\cite{adapmoe,klotski,Pre-gated,expertflow} (e.g., Fate~\cite{fate}) fetch experts only when needed, improving prefill throughput but introducing I/O bubbles during decode. Dynamic scheduling strategies~\cite{MoE-APEX,hybridmoe} (e.g., MoE-APEX~\cite{MoE-APEX}, HybriMoE~\cite{hybridmoe}) monitor runtime load and adapt expert placement at each layer, but they react to the current layer rather than anticipating the next. As Fig.~\ref{figure1} shows, none of these phase-blind approaches escapes the seesaw: on all evaluated models, the on-demand and dynamic schedulers that lead prefill throughput fall behind in decode, while static placement shows the opposite. A common limitation underlies this dilemma: scheduling decisions are made solely from current runtime state, without exploiting the inherent predictability of expert routing. Without lookahead, prefill schedulers cannot distinguish between hot experts that will be needed soon and those that will not, leading to underfetching of critical experts; decode schedulers cannot avoid unnecessary transfers that do not contribute to future tokens, leading to overfetching and I/O bubbles. This myopia prevents existing systems from breaking the seesaw trade-off.


In contrast, we observe that the two phases are predictable in complementary ways. In prefill, the parallel processing of many tokens provides sufficient context to accurately predict upcoming expert activations across layers; in decode, per-token generation follows consistent patterns that allow for stable I/O planning. Phase heterogeneity is therefore not a disturbance to be smoothed over, but a schedulable resource: instead of forcing both phases into a single scheduling framework, each phase can be modeled, scheduled, and optimized independently at the prediction, scheduling, and execution layers, decomposing the conflicting resource requirements into two separate optimization spaces.

We reframe efficient MoE inference as a phase-adaptive scheduling problem: not how to build a fixed execution flow, but how to dynamically configure its optimal operating point for each phase using predicted routing information. This problem is most pronounced in single-batch scenarios on single CPU-GPU nodes, common in interactive applications like local coding assistants or document processing tools. Here, requests arrive in single or few batches, eliminating inter-request parallelism that could mask resource contention between prefill and decode. Without such parallelism, all phase-agnostic systems face the same dilemma---using a unified scheduling strategy to adapt to two diametrically opposed optimization goals, leading to severe resource contention and pipeline stalls. Optimizing for this worst-case scenario unlocks the potential of local MoE deployment; improvements for low-batch phase conflicts naturally extend to multi-batch settings where phase conflicts are less severe~(\S\ref{subsec:scalability}).

Building on this insight, we present PDScope\footnote{Code available at \url{https://github.com/nezha-Anonymous/PDScope}.}---an inference system that treats phase-awareness as a first-class scheduling primitive. Its design includes three key elements that work in concert:

\textbf{Phase-aware prediction}, PhasePer, formalizes expert activation prediction as a phase-specific ranking problem. Exploiting the layer-group effect (\S\ref{sec:31}), it achieves 94\% Top-6 accuracy on DeepSeek-V2-Lite~\cite{deepseekv2} and Qwen3-30B-A3B~\cite{Qwen3}, providing reliable lookahead for phase-appropriate prefetching.
\textbf{Phase-adaptive scheduling}, AdaptSched, co-optimizes current layer placement and upcoming layer prefetching through cross-layer cost modeling, adopting aggressive policies during prefill and conservative policies during decode.
\textbf{Overhead-free execution engine} compiles scheduling decisions into efficient sequences through static computation graph, cache prefetching, and zero-copy I/O co-design, preventing runtime overhead from offsetting phase-aware gains.



As shown in Fig.~\ref{figure1b}, PDScope instead allocates more computation to the GPU during prefill and minimizes PCIe overhead during decode, achieving more balanced resource utilization than phase-blind schedulers. These gains translate into 2.5–10.3× higher prefill throughput and 2.1–6.1× higher decode throughput than existing systems on mainstream MoE models.


\section{Background and Related Work }
\label{II}

\subsection{Inference of MoE Models}
\label{21}

A MoE model consists of stacked layers. For each input token $x$, a layer processes it as follows: the attention module first outputs a hidden state $a$. The gating module then computes routing weights $w_i$ and selects the top-\(k\) activated experts ${e}_i$ to process $a$, yielding the layer output $x'$, which serves as input to the next layer. Formally, this can be expressed as:
\[
a = \text{Attention}(x) \quad e_i, w_i = \text{Gate}(a) \quad x' = \sum_{i=0}^{k-1} w_i \cdot \text{Expert}_i(a)
\]

A key observation is that once $a$, ${e}_i$ and $w_i$ are determined, subsequent expert selection becomes deterministic, governed by the model's fixed parameters. This establishes the theoretical foundation for expert predictability.

\begin{figure}[t]
  \centering
  \begin{subfigure}[ht]{\linewidth}
    \includegraphics[width=\linewidth]{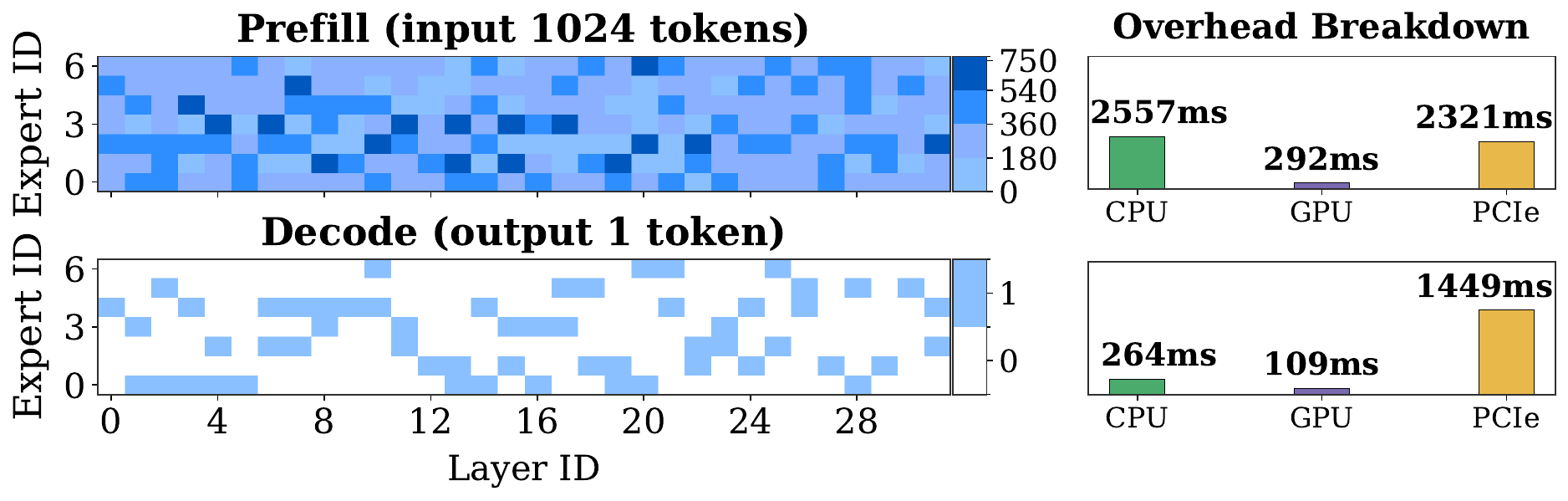}
  \end{subfigure}
  \caption{HybriMoE-scheduled Mixtral-8x7B reveals bottlenecks: prefill activates all 8 experts/layer with CPU-dominated overhead, while decode activates 2 experts/layer with PCIe-dominated latency.}
     \label{figure2} 
\end{figure}

\subsection{Disparity between Prefill and Decode}
\label{sec22}

The resource demand disparity between prefill and decode stems from their fundamentally different token-level parallelism. Fig.~\ref{figure2} quantifies this using Mixtral-8x7B on the A100 platform under a CPU-GPU load-balancing scheduler (HybriMoE~\cite{hybridmoe}).
Prefill activates 8 experts per layer, with high token-level parallelism saturating the CPU. Processing more tokens on the GPU alleviates this pressure, yet PCIe transfer latency can stall the GPU. Prefill should therefore prioritize hot experts under PCIe constraints. Decode activates only 2 experts per token; with minimal per-expert computation, loading for load balancing risks making PCIe the bottleneck. Decode should therefore minimize unnecessary I/O. These are two fundamentally different objectives.

\subsection{CPU-GPU Inference Related Work}
\label{sec:23}
Existing systems improve MoE inference through two complementary techniques: expert prefetching and expert scheduling.
\textbf{Prefetching} transfers experts in advance to overlap computation with I/O~\cite{toward,fMoE,exploiting}. Its effectiveness hinges on predicting next-layer expert activation. Existing methods fall into three categories: statistics-based methods~\cite{klotski,diff,prediction,popfetcher} assume frequently activated experts remain frequent; gate-based methods~\cite{hybridmoe,InfiniGen,ProMoE,fate} exploit correlations between adjacent-layer gating inputs; token-based methods~\cite{sida,OpenMoE,expertflow,CasMoE} reuse historical routing decisions for similar token representations.
\textbf{Scheduling} decides where to place experts and when to prefetch~\cite{Edgellm,specexec,flexgen,tang2024hobbit,MoE-GEN}. Strategies fall into three categories: static placement~\cite{ktransformers,llamacpp,sglang,vllm} avoids runtime I/O entirely; on-demand loading~\cite{lina,expertflow,Moelightning,EdgeMoE} keeps experts in CPU memory and loads them only when needed; dynamic scheduling~\cite{fiddler,hybridmoe,TriMoE} monitors hardware load and dispatches experts to idle devices, though this introduces inspection overhead.

\begin{figure}[t]
  \centering
    \includegraphics[width=\linewidth]{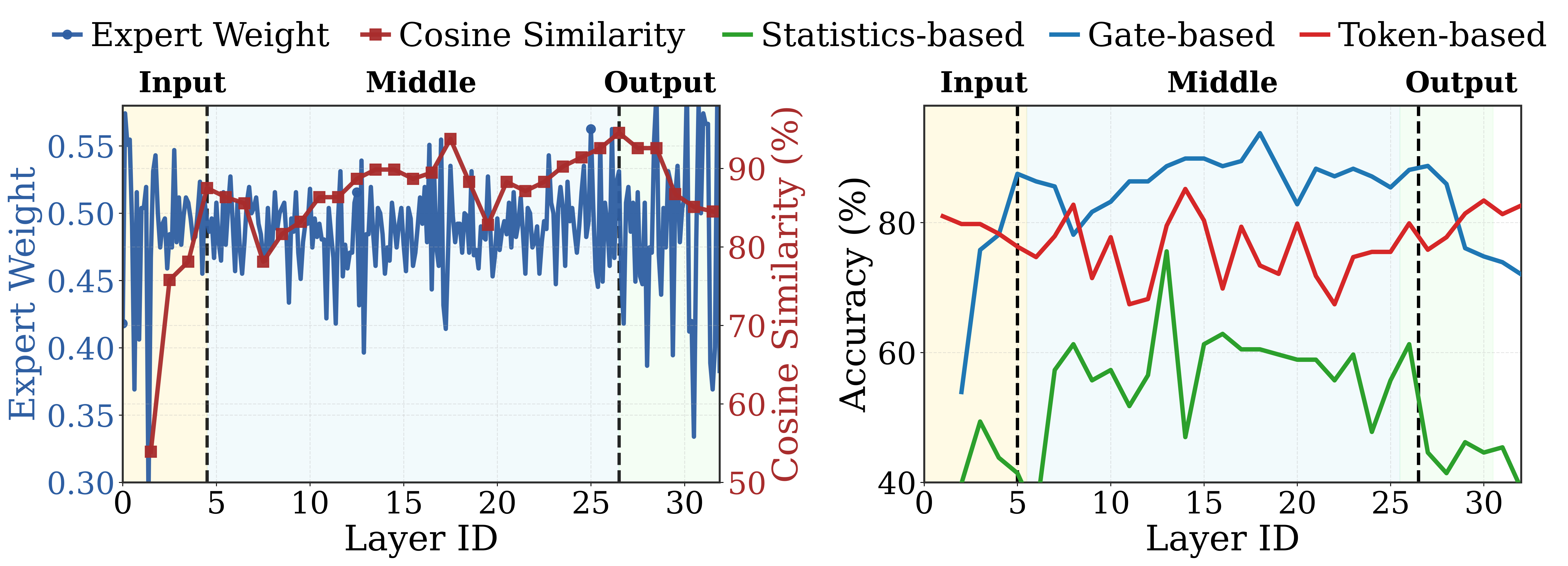}
  \caption{ Linking group-wise patterns in routing weight and inter-layer similarity to prediction methods in Mixtral-8x7B.}
     \label{figure4} 
\end{figure}

While these methods optimize across the spatial dimension (where to place) and the temporal dimension (when to prefetch), they remain agnostic to the phase dimension (prefill vs. decode). Consequently, their unified logic forces a single operating point that statically compromises between compute-bound and I/O-bound regimes. PDScope distinguishes itself by explicitly exposing the phase axis as a first-class scheduling dimension, enabling two distinct operating points that are independently optimized.

\section{Challenges and Motivation}

\subsection{Expert Prefetching Challenge}
\label{sec:31}

Analysis of Mixtral-8x7B~\cite{moe} reveals a layer-group effect\footnote{\textit{Appendix} \S\ref{look} observes similar patterns in other models.} (Fig.~\ref{figure4}), where the distributions of routing weights $w_i$ and the similarities of hidden states $a$ across layers cluster MoE layers into input, middle, and output groups with distinct statistical characteristics. As a result, no single heuristic works globally: statistics-based methods suffer from suboptimal overall accuracy; token-based methods falter in the middle groups; gate-based methods, while effective in the middle, fail where inter-layer similarity is low. Exploiting expert activation predictability thus requires layer-group awareness—yet manual or statistical identification of each group's properties is challenging. A learnable predictor, using relevant variables as input features to capture these associations in a data-driven manner, offers a more viable solution.

\subsection{Expert Scheduling Challenge}

\label{sec:32}

\begin{figure}[t!]
  \centering
  \begin{subfigure}[ht]{\linewidth}
    \includegraphics[width=\linewidth]{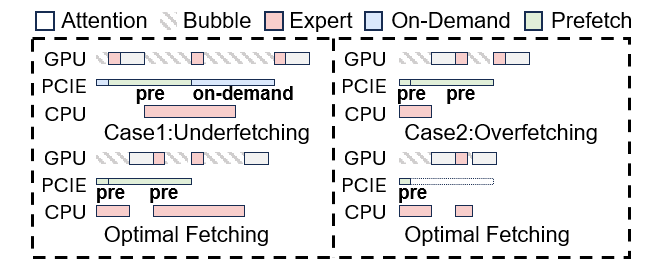}
  \end{subfigure}
\caption{Underfetching in prefill and overfetching in decode caused by layer-by-layer greedy scheduling.}
     \label{figure5} 
\end{figure}

Existing scheduling strategies attempt to handle the prefill/decode disparity with unified logic. As shown in Fig.~\ref{figure1}, this seesaw effect is stark: static placement excels in decode but fails in prefill; on-demand loading and dynamic scheduling show the opposite. This section examines dynamic scheduling's failure mechanism: it adapts via real-time load balancing, appearing to be the most promising approach, yet still falls short due to its layer-by-layer greedy nature. Taking HybriMoE~\cite{hybridmoe} as an example, the scheduler monitors per-layer load and greedily assigns hot experts to idle GPUs and cold experts to idle CPUs, pursuing instant balance. This yields suboptimal results in both phases (Fig.~\ref{figure5}):

\textbf{Case 1 (prefill):} When only cold experts remain, the scheduler still loads them to the GPU, consuming PCIe bandwidth that should go to prefetching the next layer's hot experts, while the cold experts could execute on CPU without PCIe contention. The current layer achieves local optimum, but the next layer stalls-degrading global efficiency.
\textbf{Case 2 (decode):} When no experts remain in the current layer, the scheduler triggers prefetching for the next layer. If this prefetching does not complete before the next layer's attention and gating finish, it introduces I/O-induced bubbles, exacerbating the decode-phase PCIe bottleneck.

Prefill thus requires aggressive prefetching to maximize  tokens processed on the GPU, while decode must avoid unnecessary I/O. Reconciling these conflicting demands requires moving beyond layer-by-layer greediness toward cross-layer coordination that adapts scheduling strategies to each phase.

\begin{table}[t!]
\centering
\small
\caption{PCIe 4.0x16 single-expert transfer latencies (ms).}
\label{tab:expert_transfer_overhead}
\begin{tabular}{@{}lcccc@{}}
\toprule
\textbf{Model} & \textbf{Benchmark} & \textbf{Prefill} & \textbf{Decode} & \textbf{Theoretical} \\
\midrule
DeepSeek~\cite{deepseekv2} & 1.95 & 2.78 & 2.06 & 0.53 \\
Qwen~\cite{Qwen3} & 1.34 & 1.88 & 1.38 & 0.29 \\
Mixtral~\cite{moe} & 33.5 & 51.1 & 33.8 & 10.67 \\
\bottomrule
\end{tabular}
\\[2pt]
\parbox{\linewidth}{\footnotesize Benchmark: micro-benchmark without CPU contention; Prefill/Decode: measured during inference with resource competition; Theoretical: minimum latency from PCIe bandwidth limit.}
\label{tab3}
\end{table}

\subsection{Execution Challenge: Overhead Sensitivity}
\label{sec:33}

Scheduling gains are bounded by system overhead. Executing a single expert incurs three non-computational costs: CPU cache loading, PCIe transfer, and GPU kernel launch. These overheads differ between prefill and decode, so they cannot be used indiscriminately as scheduling criteria.

In prefill, the CPU is busy, and I/O preparation steps (e.g., pinned memory allocation and DMA setup) compete with computation for CPU resources. Table~\ref{tab3} shows that prefill transfer latency is significantly amplified vs. benchmarks (e.g., DeepSeek-V2-Lite: 1.95ms vs. 2.78ms); the scheduler's real-time advantage is eroded by resource contention.

Decode faces a different predicament: low token-level parallelism prevents overhead amortization. Table~\ref{tab2} reveals a significant gap between computation time and end-to-end runtime for expert computation: overheads from cache loading, kernel launches, and hardware status polling are substantial, particularly on CPU (e.g., Mixtral-8x7B on Intel CPU: 2.88ms compute vs. 4.25ms runtime). Any scheduling query, status polling, or extra transfer directly contributes to end-to-end latency instead of being masked by computation.

Therefore, scheduling decisions must account for phase-specific overhead characteristics, ensuring that their gains exceed the overhead they introduce. At the same time, the scheduler itself should minimize its own pipeline burden.

\begin{table}[t]
\centering
\small
\caption{ Single-expert, single-token computation time  and end-to-end runtime  across hardware platforms (ms).}
\resizebox{\linewidth}{!}{%
\begin{tabular}{@{}lcccccc@{}}
\toprule
 & \multicolumn{2}{c}{\textbf{Intel 6230R CPU}} & \multicolumn{2}{c}{\textbf{AMD EPYC CPU}} & \multicolumn{2}{c}{\textbf{RTX 3090 GPU}} \\
\cmidrule(r){2-3} \cmidrule(lr){4-5} \cmidrule(l){6-7}
\textbf{Model} & \textbf{Compute} & \textbf{Runtime} & \textbf{Compute} & \textbf{Runtime} & \textbf{Compute} & \textbf{Runtime} \\
\midrule
DeepSeek~\cite{deepseekv2} & 0.21 & 0.43& 1.5  & 2.1 & 0.17 & 0.29 \\
Qwen~\cite{Qwen3} & 0.16 & 0.34& 1.2  & 1.64 & 0.13 & 0.24 \\
Mixtral~\cite{moe} & 2.88 & 4.25& 9.8  & 11.2 & 0.33 & 0.44 \\
\bottomrule
\end{tabular}%
}
\label{tab2}
\end{table}

\section{System Design}

\subsection{Overview}
\label{sec:overview}

\begin{figure}[t]
  \centering
  \begin{subfigure}[ht]{\linewidth}
    \includegraphics[width=\linewidth]{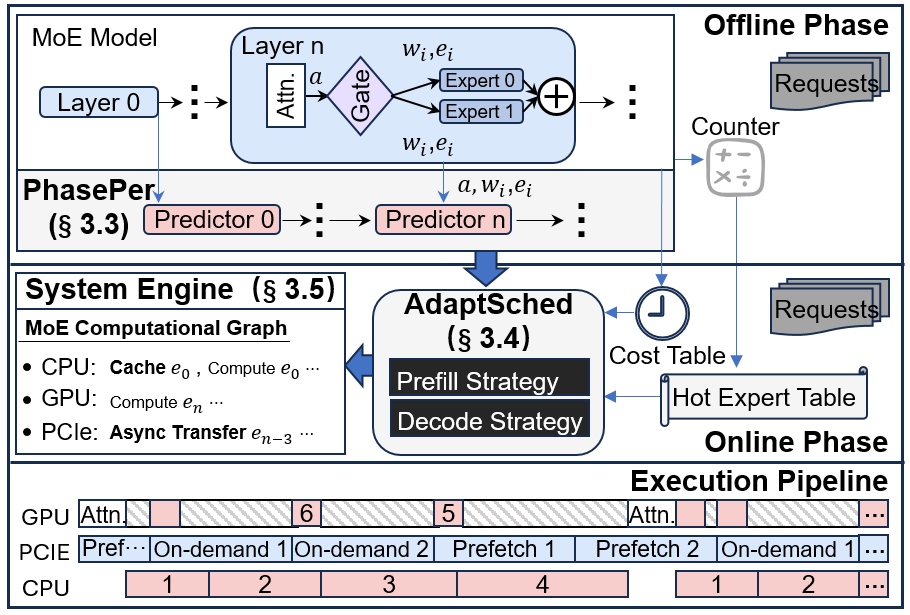}
  \end{subfigure}
  \caption{System overview of PDScope.}
     \label{figure7} 
\end{figure}

PDScope is a phase-aware inference system for memory-constrained MoE inference. As shown in Fig.~\ref{figure7}, its workflow comprises offline profiling and online execution.

\textbf{Offline phase.} The system profiles the MoE model, collecting per-layer traces (hidden states, routing weights, expert IDs) to train PhasePer. It also builds a cost table mapping token count to runtime cost on CPU and GPU, which underpins AdaptSched's global optimization and adapts to the target platform's bandwidth and compute capability. 
 Hot experts are recorded in a table and preloaded into GPU memory as a static cache to amortize cold-start overhead.


\textbf{Online phase.} PhasePer provides predictions using phase-specific models. AdaptSched consumes these predictions to co-optimize current-layer placement and upcoming-layer prefetching: aggressive prefetching in prefill sustains GPU throughput; conservative I/O in decode minimizes PCIe bubbles. Decisions are compiled into a static computation graph that prescribes all operations and data movements, eliminating per-expert runtime scheduling decisions. Execution follows this graph: a graph-guided cache prefetcher hides memory latency, while a zero-copy asynchronous I/O engine maximizes PCIe bandwidth via pipelined transfers. 
This runtime stack directly addresses the phase-sensitive overheads analyzed in \S\ref{sec:33}, ensuring scheduling decisions translate to performance without adding runtime overhead that would offset gains.

\subsection{Learnable Phase-Aware Predictor}

\subsubsection{Design principle}

PhasePer builds on the deterministic mapping from $(a, w_i, {e}_i)$ to the next layer's expert activation ${e}_i'$ established in \S\ref{21}. It provides separate dedicated predictors for prefill and decode, allowing each to specialize in the statistical patterns of its respective phase.

\subsubsection{Network architecture}

PhasePer learns the cross-layer function ${e'}_i=\text{PhasePer}_{\theta}(a, w_i, {e}_i)$, mapping the current layer’s full state to the next layer’s expert activation. This rich input context enables it to capture data-driven selection patterns, yielding more accurate predictions than methods that simply approximate the next layer’s gate.

To match the layer-group effect described in \S\ref{sec:31}, PhasePer instantiates a dedicated lightweight predictor for each MoE layer. Predictors for layers in the input and output groups, where activation is skewed and features are noisier, use shallow networks to prevent overfitting. For the middle group, where inter-layer features are stable, predictors are deeper and include a residual block modulated by a gating unit that takes $a$ as input to dynamically calibrate the network. 


\subsubsection{Training strategy}
We collect training data by performing offline inference on 128 ShareGPT~\cite{sharegptv3} batches, each containing a 4096-token prompt followed by 512 generated tokens. Per-layer records ($a$, $w_i$, ${e}_i$) are organized separately by phase, prefill and decode, to form two dedicated training sets. This phase-isolated data collection ensures that each predictor learns from phase-specific activation patterns. Cross-dataset generalization is evaluated in \S\ref{preacc}.

\noindent\textbf{Loss function.} Loss functions are phase-customized. For input/output groups in prefill, a hybrid loss counteracts severe token-level class imbalance and emphasizes hard examples:
\begin{equation}
\begin{aligned}
&\mathcal{L}_{\text{prefill}} = \mathcal{L}_{\text{expert}} + \lambda\mathcal{L}_{\text{focal}}\\
\mathcal{L}_{\text{expert}}
  &= -\frac{1}{N}\sum\nolimits_{i=1}^{N}
     \frac{1}{f_i}\,\text{BCE}(y_i,p_i)\\
\mathcal{L}_{\text{focal}}
  &= -\frac{1}{N}\sum\nolimits_{i=1}^{N}
     (1-p_i^{y_i})^{\gamma}\,\text{BCE}(y_i,p_i)
\end{aligned}
\end{equation}

Here, $f_i$ denotes activation frequency of expert $i$; $y_i\in\{0,1\}$ indicates ground-truth activation; $p_i$ is the predicted probability; $N$ is number of experts;  BCE is the binary cross-entropy loss; and $\lambda,\gamma\ge 0$ are layer-specific hyper-parameters.

During  decode, the expert imbalance is negligible as tokens are processed one by one. Consequently, the loss for the input and output groups reduces to the focal term alone:
\begin{equation}
\mathcal{L}_{\text{decode}} = \mathcal{L}_{\text{focal}}
\end{equation}


Middle-group predictors in both phases use standard BCE for stable gradients and faster convergence.

\noindent\textbf{Online fine-tuning.}
PhasePer includes a lightweight online learner. It monitors prediction error on a sliding window of 16 input-group tokens. If the average error exceeds a preset threshold (25\% in our implementation), a fine-tuning cycle is triggered using runtime data from 128 prefill and 32 decode tokens, completing within 2.1 ms.

\begin{figure}[t]
  \centering
  \begin{subfigure}[ht]{\linewidth}
    \includegraphics[width=\linewidth]{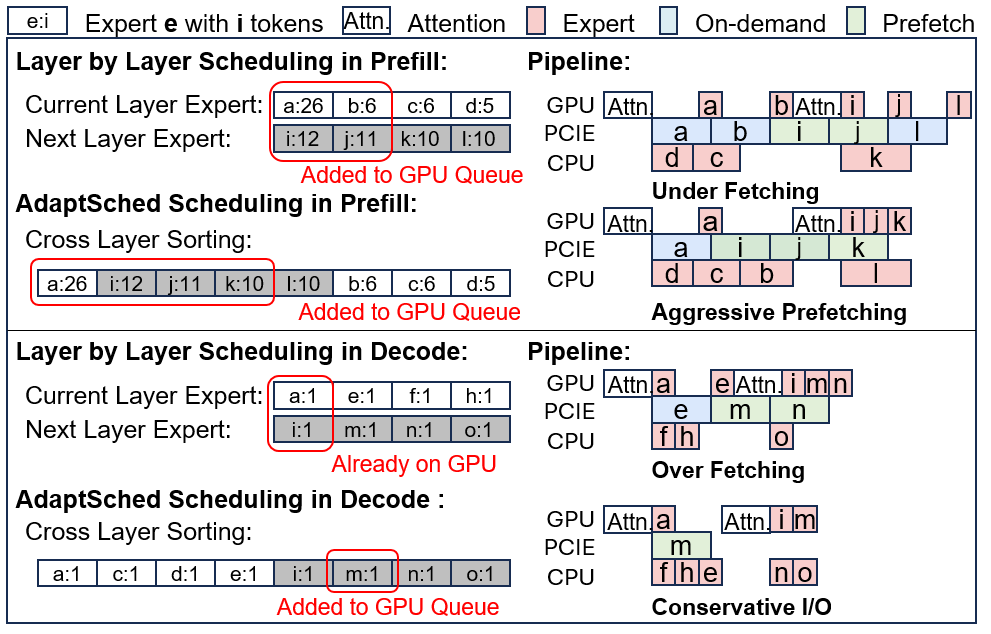}

  \end{subfigure}
\caption{AdaptSched replaces layer-by-layer greediness: prefill eliminates underfetching via aggressive prefetching; decode eliminates overfetching via conservative I/O.}
     \label{figure9}  
\end{figure}

\subsection{Phase-Adaptive Cross-layer Scheduler}

\subsubsection{Design principle}

As shown in Fig.~\ref{figure9}, AdaptSched differs from layer-by-layer schedulers by leveraging the predictions of PhasePer for cross-layer visibility\footnote{Single-layer lookahead is sufficient; see \textit{Appendix} \S\ref{}.}. This enables it to co-optimize current-layer placement and next-layer prefetching to maximize PCIe utilization. In prefill, it prioritizes token-heavy experts to maximize  tokens processed on the GPU. In decode, it balances on-demand loading, prefetching, and unloading to avoid I/O inefficiencies. Prefetch depth is dynamically calibrated based on predictor confidence to avoid interfering with critical on-demand traffic.

\begin{figure}[t]
  \centering
  \begin{subfigure}[ht]{\linewidth}
    \includegraphics[width=\linewidth]{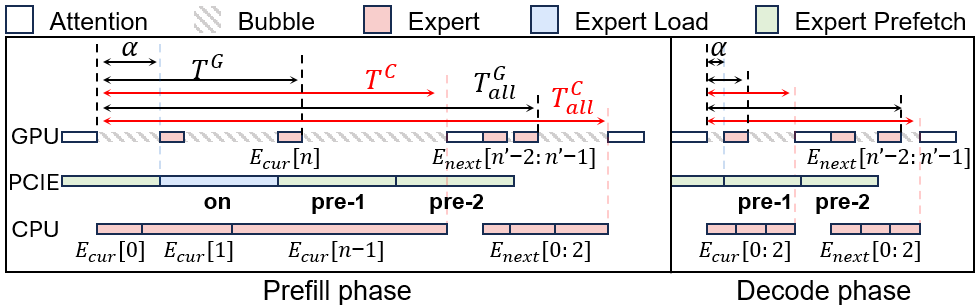}

  \end{subfigure}
\caption{Phase-specific cost model of AdaptSched.}
     \label{figure99}  
\end{figure}

\subsubsection{Problem modeling}
\label{subsec:formal-model}

The scheduler rests on three premises. 
\textbf{(1) Cross-layer sorting:} After excluding experts already on the GPU or being loaded, lists $E_{cur}[0\ldots n]$ and $E_{next}[0\ldots n'-1]$ are formed by sorting experts in ascending order of pending tokens. 
\textbf{(2) On-demand monotonicity:} To maximize I/O benefit, loading decisions are monotonic, if $E_{cur}[i]$ requires loading, so must hotter experts $E_{cur}[i+1: n]$.
\textbf{(3) Serial I/O:} Transfers are executed sequentially, as single-expert transfers already reach 66--80\% PCIe peak (Table~\ref{tab4}); adding parallelism risks congestion without faster arrival.

\subsubsection{Prefill scheduling}
\label{secschel}
The scheduler operates in three steps: Step 1 establishes a global resource allocation direction; Step 2 determines the current layer's on-demand loading set; and Step 3 decides the prefetch set.

\noindent\textbf{Step 1: Global queue construction via cross-layer cost modeling.}
Algorithm~\ref{alg:AdaptSched} merges and sorts experts from the current layer $E_{cur}$ and the predicted next layer $E_{next}$ in ascending order of token count, forming $E_{all}$ in line 9. For the expert at position $i$ in $E_{all}$, the scheduler evaluates a cross-layer cost model, comparing two completion times depicted in the prefill timeline of Fig.~\ref{figure99}:

\begin{equation}
\begin{aligned}
T^G_{all} &= \alpha + (n + n' - i + 1)\cdot t_{io} + t_g \\
T^C_{all}
&= t_c(E_{all}[0: i]) + t_{attn}
\label{eq8}
\end{aligned}
\end{equation}
Here, $\alpha$ is I/O startup latency, $t_{io}$ and $t_g$ are per-expert transfer and GPU computation times, and $t_c(\cdot)$ computes CPU execution time for a set of experts. $T^G_{all}$ estimates the cost if experts $E_{all}[i: n+n']$ are loaded and executed on the GPU, while $T^C_{all}$ estimates the cost of executing experts $E_{all}[0: i]$ on the CPU (plus attention latency $t_{attn}$). If $T^G_{all} < T^C_{all}$, experts $E_{all}[i: n+n']$ join the global queue $L_{global}$. This step, implemented in lines 10-12, selects experts that benefit most from GPU execution from a cross-layer perspective.

\noindent\textbf{Step 2: Layer-local re-evaluation for on-demand loading.}
Since $L_{global}$ interleaves experts from both layers, directly following it may imbalance the current layer, e.g., all current-layer experts could be assigned to $L_{global}$, leaving the CPU idle. This can occur due to cross-layer distribution differences or uncertainties in PhasePer’s predictions. To correct this, AdaptSched re-evaluates the subsequence of experts belonging to the current layer within $L_{global}$. For a candidate boundary index $i'$ in $E_{cur}$, it compares:

\begin{equation}
\begin{aligned}
T_G = \alpha + (n - i' + 1)\cdot t_{io}  + t_g   \ \ 
T_C = t_c(E_{cur}[0: i'-1])
\end{aligned}
\end{equation}
The smallest $i'$ with $T_G < T_C$ finalizes the set $L_{on}$ of experts to be loaded on demand. This step, implemented in lines 14-15, guarantees load balance within the current layer.

\begin{algorithm}[t]
\fontsize{8.5pt}{10.5pt}\selectfont
\caption{AdaptSched Scheduling for Layer $l$}
\label{alg:AdaptSched}
\LinesNumbered          
\KwIn{Phase $\phi \in \{\text{prefill}, \text{decode}\}$, lists $E_{cur}[0\ldots n]$ and $E_{next}[0\ldots n'-1]$, balance target $n_c$ }
\KwOut{Experts to load $L_{on}$, prefetch $L_{pre}$ and unload $L_{un}$}
\If{$\phi = \text{decode}$}{
    \uIf{$n> n_c$ and $n'-1 \leq n_c$}{
        \Return $(L_{on},\emptyset, \emptyset)$\tcp*{\fontsize{8pt}{9pt}\selectfont Mode a:Load focus}
    }
    \uElseIf{$n \leq n_c$ and $n'-1 > n_c$}{

        \Return $(\emptyset, L_{pre}, L_{un})$\tcp*{\fontsize{8pt}{9pt}\selectfont Mode b:Prefetch focus}
    }
    \uElseIf{$n\leq n_c$ and $n'-1 \leq n_c$}{
        \Return $(\emptyset,\emptyset, L_{un})$\tcp*{\fontsize{8pt}{9pt}\selectfont Mode c:Unload focus}
    }    
    
    \tcc{\fontsize{8pt}{9pt}\selectfont Mode d or prefill: Cross-layer logic}
}
\textbf{Step 1: Build cross-layer global queue:}\;
$E_{all} \gets \textsc{MergeAndSortByToken}(E_{cur} ,  E_{next})$\;
\For{$i \gets 0$ \KwTo $|E_{all}|-1$}{
    \If{\fontsize{8pt}{9pt}\selectfont $\textsc{GlobalGpuCost}(i) < \textsc{GlobalCpuCost}(i)$}{
        $L_{global} \gets E_{all}[i:]$; \textbf{Break}\;
    }
}
\textbf{Step 2: Determine on-demand loading boundary:}\;
$i' \gets \textsc{FindSmallestIndex}(L_{global} \cap E_{cur})$ such that $\textsc{LocalGpuCost}(i') < \textsc{LocalCpuCost}(i')$ \;
$L_{on} \gets E_{cur}[i':]$\;
\textbf{Step 3: Decide prefetches:}\;
$T_{gap} \gets \textsc{ComputeCpuBubble}(i')$\;
$f \gets \lfloor (T_{gap} + t_{attn}) / t_{io} \rfloor$\;
\If{$\textsc{PrefetchBenefit}(f) > 0$}{
    $L_{pre} \gets \text{top-$f$ from } E_{next} $\;
}
\Else 
{
    $L_{pre} \gets \text{top-$(f-1)$ from } E_{next} $\;
}
\Return $(L_{on}, L_{pre}, \emptyset)$

\end{algorithm}

\noindent\textbf{Step 3: Confidence-aware prefetching via expected utility.}
After $L_{on}$ is set, remaining current-layer experts run on CPU. The gap $T_{gap} = T_C - T_G$ represents an I/O bubble, which plus $t_{attn}$ determines maximum hidden prefetches: $f = (T_{gap} + t_{attn}) / t_{io}$. Issuing the $|f|$-th prefetch depends on a confidence-aware benefit criterion $\xi$ which computes the expected utility of prefetching by incorporating the predictor’s hit rate $R_{hit}$ for high-priority experts:
\[
\xi = R_{\text{hit}} \cdot (f - |f| + 1) \cdot t_{{io}} - (1 - R_{\text{hit}}) \cdot (|f| - f) \cdot t_{{io}}
\]
The first term is the expected gain (weighted by hit rate), and the second term is the expected loss (weighted by miss rate). Prefetching proceeds only if the net expected utility $\xi > 0$. This logic, encapsulated in the \textsc{PrefetchBenefit} function, determines the final prefetch set $L_{pre}$ (lines 19-22).

\subsubsection{Decode scheduling} 
\label{decsche}

In decode, each token activates top-$k$ experts per layer, making load regular. AdaptSched employs a conservative I/O strategy: it coordinates on-demand loading, prefetching, and unloading based on deviations from the ideal load-balanced state. The optimal CPU-resident expert count $n_c$ minimizes makespan: $n_c = \arg\min \max((k - n_c) \cdot t_g, n_c \cdot t_c)$. This drives four modes (lines 1-7):

\textbf{Mode a: load focus:} If the current layer's CPU-resident expert count is above $n_c$ while the next layer's count is at or below this threshold, the scheduler prioritizes on-demand loading for the current layer, reusing Step~2 logic of Algorithm~\ref{alg:AdaptSched} with a revised cost model:
\begin{equation}
T_G = \alpha + (n - i' + 1)\cdot t_{io} + t_g \quad
T_C = t_c(E_{cur}[0: i'])
\end{equation}
Unlike the prefill formulation (which keeps boundary experts on CPU to reserve PCIe bandwidth for prefetching), this revised model minimizes current-layer end-to-end latency.

\textbf{Mode b: prefetch focus.} If the current layer's CPU-resident count falls below $n_c$ while next layer's count is above, the scheduler unloads redundant experts and uses PCIe bandwidth for prefetching. Prefetching reuses Step~3 logic of Algorithm~\ref{alg:AdaptSched}, subject to an additional constraint:
\begin{equation}
(|f| - f)\cdot t_{io} + t_g < t_c(E_{next}[0: n'-|f|])
\end{equation}
This prevents prefetching from causing next-layer GPU completion time to exceed its CPU-only completion time without boundary prefetching, avoiding decode over-prefetching.

\textbf{Mode c: unload focus:} If both current and next layers' CPU-resident counts below $n_c$, the scheduler only unloads (backups of all experts are on CPU, so no I/O overhead).

\textbf{Mode d: cross-layer focus:} If both counts are above $n_c$, the scheduler reuses the full cross-layer logic in Step~1--3,  with prefetching subject to the constraint from Mode (b).

\subsection{System-Level Optimization Engine}

The system engine translates AdaptSched's scheduling decisions into performance by co-optimizing computation orchestration, memory access, and data movement.

\subsubsection{Static graph orchestration and runtime optimization}
\label{graph}

AdaptSched's decisions are compiled into a static computation graph for each layer through a two-phase strategy that decouples scheduling from execution.

\textbf{Pre-compilation phase.} While the runtime executes layer $l-1$, the scheduler uses PhasePer's predictions to pre-compile a draft graph for layer $l$. This draft prescribes placements and prefetches only for experts with high prediction confidence—top-ranked (hot) and bottom-ranked (cold) experts in $E_{next}$—decoupling their execution from runtime scheduling.
\textbf{Recompilation phase.} Once the true gating outputs of layer $l$ become available, PDScope recompiles the complete static graph, resolving all remaining experts using the full AdaptSched logic. The recompiled graph overwrites decisions from the draft, ensuring that prediction errors never propagate to execution.

This pipeline rolls forward: while layer $l$ executes, the scheduler pre-compiles a draft for layer $l+1$. Pre-compilation overlaps scheduling and graph construction overhead with computation, while recompilation corrects for prediction uncertainty using ground-truth information. The runtime then executes each static graph asynchronously. The graph further orchestrates two classes of execution optimizations: CPU-side prefetching of predicted expert weights into the last-level cache before computation begins, and GPU-side kernel launch batching that consolidates all resident and prefetched experts into a single coordinated submission between the attention and MoE phases. This design eliminates per-expert dynamic decision overhead while also reducing the execution latency of expert computation.



\subsubsection{Zero-copy asynchronous I/O optimizer}
\label{secio}

The AsyncIO optimizer pre-registers two types of expert weight buffers in GPU memory: a ping-pong buffer for on-demand loading and multiple buffers for prefetching, turning costly dynamic registration into a one-time overhead. This design ensures prefetching for layer $l+1$ never conflicts with on-demand transfers for layer $l$. To maximize bandwidth, AsyncIO transfers the three matrices of each expert (\texttt{gate\_proj, up\_proj, down\_proj}) concurrently via three independent CUDA streams. All transfers use pinned host memory and asynchronous copies, minimizing CPU intervention. As shown in Table~\ref{tab4}, AsyncIO achieves 66–80\% PCIe bandwidth utilization, approaching the practical peak.

\begin{table}[t!]
\centering
\small
\caption{Expert transfer performance and PCIe bandwidth utilization after AsyncIO optimization in the prefill phase.}
\label{tab:asyncio_optimization}
\resizebox{\linewidth}{!}{%
\begin{tabular}{@{}lcccccc@{}}
\toprule
& \multicolumn{2}{c}{\textbf{DeepSeek}} & \multicolumn{2}{c}{\textbf{Qwen}} & \multicolumn{2}{c}{\textbf{Mixtral}} \\
\cmidrule(lr){2-3} \cmidrule(lr){4-5} \cmidrule(l){6-7}
 & Cost (ms) & BW Util. & Cost (ms) & BW Util. & Cost (ms) & BW Util. \\
\midrule
PCIe 3.0 ×16 & 1.39 & 76\% & 0.79 & 72\% & 26.7 & 80\% \\
PCIe 4.0 ×16 & 0.76 & 70\% & 0.44 & 66\% & 13.98 & 76\% \\
\bottomrule
\end{tabular}%
}
\label{tab4}
\end{table}

\begin{figure*}[t!]
  \centering
  \begin{subfigure}[ht]{\linewidth}
    \includegraphics[width=\linewidth]{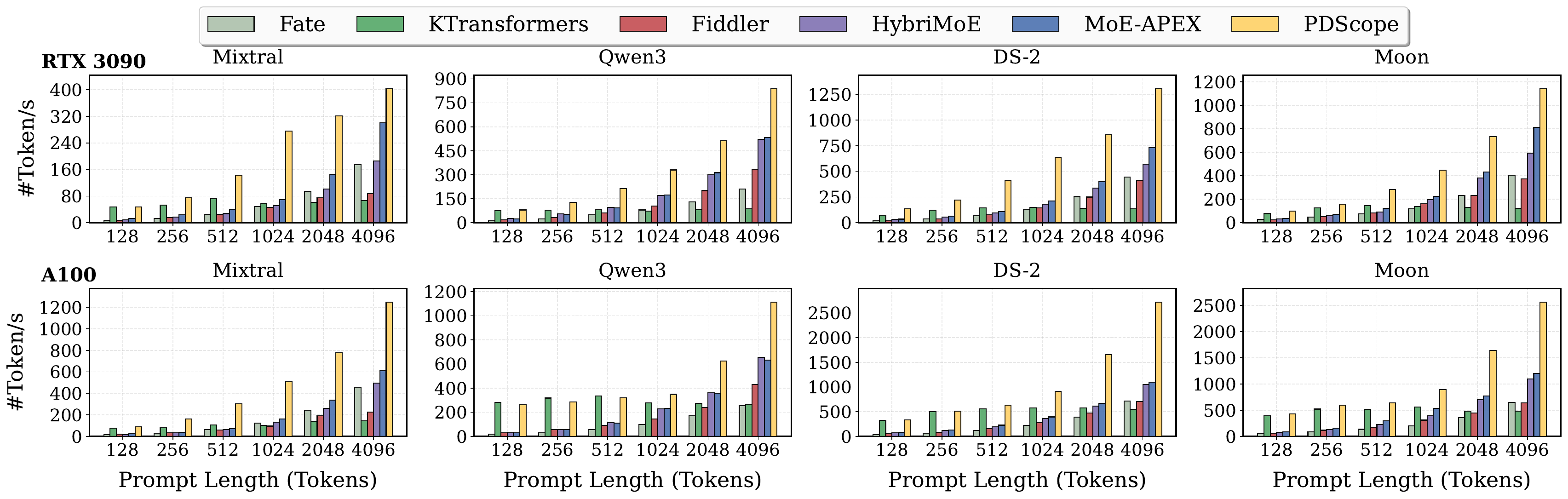}
  \end{subfigure}
  \caption{Throughput comparison of PDScope and baseline systems across models and hardware in the prefill phase.}
     \label{figure11} 
\end{figure*}

\section{Evaluation}

\subsection{Experimental Setup}

\label{expset}

\noindent\textbf{Platform.}
We evaluate on two hardware platforms: A100 (NVIDIA A100 GPU with dual-socket Intel Xeon Gold 6230R\footnote{Intel Xeon Gold 6230R supports AVX-512 but not AMX.} CPUs) and RTX 3090 (NVIDIA RTX 3090 GPU with AMD EPYC 7452 CPU, allocated 16 cores per GPU on a cloud node). Both use PCIe 4.0x16. 

\noindent\textbf{Models.}
The main evaluation uses four BF16 MoE models: Mixtral-8x7B~\cite{moe}, Qwen3-30B-A3B~\cite{Qwen3}, DeepSeek-V2-Lite~\cite{deepseekv2}, and Moonlight-16B-A3B~\cite{moonlight}. Their key differences are summarized in Table~\ref{tab5}. For larger-model scalability, we include DeepSeek-V4-Flash (INT8, 291B)~\cite{ds4} in \S\ref{subsec:scalability}.

\noindent\textbf{Baselines.}
PDScope compares against five state-of-the-art baselines spanning the three paradigms discussed in \S\ref{sec:23}: KTransformers\footnote{KTransformers v0.3.2, consistent with their paper.}~\cite{ktransformers} pre-allocates experts via static placement with customized AVX-512 kernels; Fate~\cite{fate} fetches experts on-demand using gate-based prediction; Fiddler~\cite{fiddler}, HybriMoE~\cite{hybridmoe}, and MoE-APEX~\cite{MoE-APEX} employ dynamic scheduling with runtime load monitoring, with MoE-APEX using INT4 compression selectively on less important experts. All baselines are evaluated at their individually tuned optimal configuration under the same memory constraints, after three rounds of method-specific warm-up (e.g., kernel warm-up for KTransformers, dynamic hot expert search for HybriMoE).

\noindent\textbf{Memory configuration.}
We cap GPU memory at 24GB. After allocating non-expert weights and runtime buffers, 5.8–8.7 GB remains for GPU-resident experts, accommodating each method's allocation strategy (e.g., PDScope's hot expert table, KTransformers' static placement).

\noindent\textbf{Metrics.}
Throughput is measured separately for the prefill and decode phases\footnote{We adopt the same metrics set from KTransformers~\cite{ktransformers}.}. All results are averaged over ten runs.

\noindent\textbf{Datasets.} 
We use ShareGPT~\cite{sharegptv3} with 128–4096 input tokens and 64 output tokens. For out-of-domain evaluation, PhasePer (trained on ShareGPT) is tested on LLaVA~\cite{llava} (vision-language reasoning) and DPO~\cite{DPO} (preference-tuning).

\begin{table}[t!]
\centering
\small
\caption{Configuration of evaluated MoE models.}
\resizebox{\linewidth}{!}{
\begin{tabular}{@{}lcccc@{}}
\toprule
 & \textbf{Mixtral} & \textbf{Qwen} & \textbf{DeepSeek} & \textbf{Moonlight} \\
\midrule
Layers & 32 & 48 & 26 & 26 \\
Routed Experts & 8 & 128 & 64 & 64 \\
Activated Experts & 2 & 8 & 6 & 6 \\
Expert Memory (BF16) & 336 MB & 9 MB & 16.5 MB & 16.5 MB \\
\bottomrule
\end{tabular}%
}
\label{tab5}
\end{table}

\subsection{End-to-End Performance}

\subsubsection{Prefill throughput}
\label{subsec:prefill-throughput}

Fig.~\ref{figure11} presents prefill throughput under varying input lengths across hardware platforms.

\textbf{Generalization across hardware platforms.} Under identical PCIe bandwidth, the systems exhibit distinct hardware dependencies. We compute the geometric mean of A100/RTX 3090 throughput ratios across all (model × length) configurations. The ratios for Fate, Fiddler, HybriMoE, MoE-APEX, and PDScope are approximately 1.67, 1.92, 1.85, 1.72, and 1.88, respectively, indicating similar scaling factors across platforms. In contrast, KTransformers exhibits a ratio of 3.23, highlighting its high sensitivity to CPU performance.

\textbf{Generalization across model architectures.} For Mixtral-8x7B, the relative performance of KTransformers and Fiddler declines when input length exceeds 1024 tokens. KTransformers is limited by its static strategy, while Fiddler struggles to schedule hot experts accurately, highlighting the limitations of static placement and simple dynamic scheduling under high-concurrency, compute-intensive workloads. For models with more but smaller experts, effective CPU-GPU load balancing is more critical, favoring HybriMoE, MoE-APEX, and PDScope. The main difference among these methods lies in I/O transfer efficiency and the hot/cold expert distribution of each transfer, which together determine the total token throughput the GPU can process per unit time.

\textbf{Scalability across input lengths.} KTransformers throughput peaks between 512 and 1024 tokens before declining, indicating inability to adapt to long-prompt compute intensity. Although Fiddler and HybriMoE throughput increases with sequence length, their scaling efficiency lags significantly due to lack of cross-layer coordination: longer inputs activate more experts, increasing I/O overhead, while the widening cost gap between cold and hot experts exacerbates the penalty of suboptimal placement. MoE-APEX suffers from quantization decision overhead under long sequences, resulting in a scaling curve close to that of HybriMoE. In contrast, PDScope exhibits the steepest throughput growth as input length increases, benefiting from cross-layer scheduling and systematic overhead optimization.

In summary, during prefill, PDScope achieves up to 10.3× the throughput of Fate, 8.6× of KTransformers, 5.8× of Fiddler, 4.4× of HybriMoE, and 2.5× of MoE-APEX.

\begin{figure}[t]
  \centering
  \begin{subfigure}[ht]{\linewidth}
    \includegraphics[width=\linewidth]{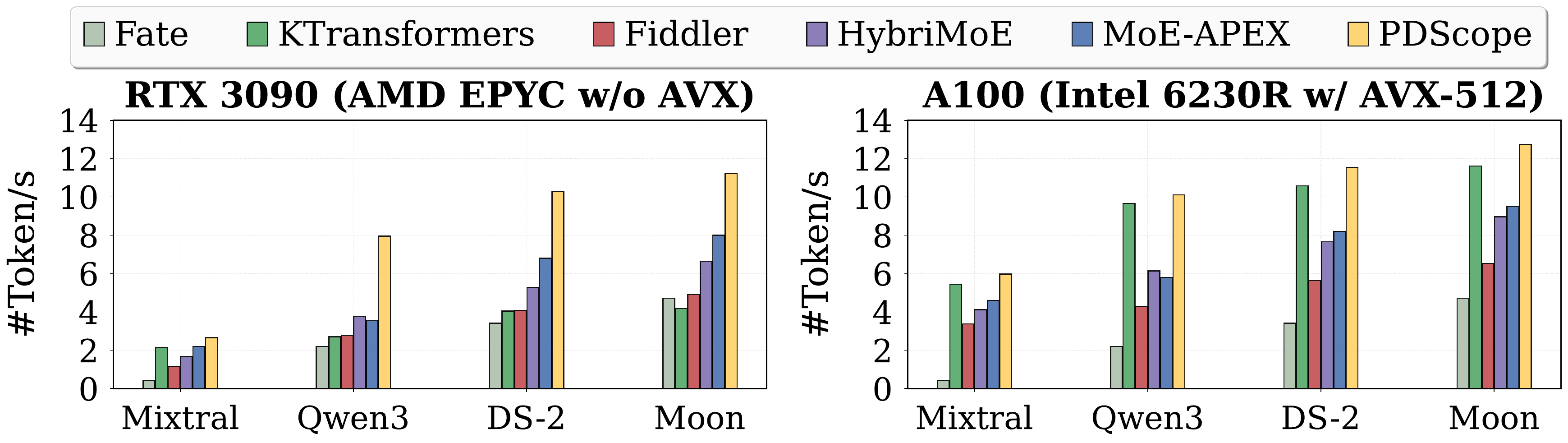}
  \end{subfigure}
  \caption{Throughput comparison of PDScope and baseline systems in the decode phase.}
     \label{figure12} 
\end{figure}

\subsubsection{Decode throughput}
\label{subsec:decode-throughput}

Fig.~\ref{figure12} presents decode throughput across two hardware platforms. The cross-platform geometric mean ratios are: Fate 1.71×, Fiddler 1.69×, HybriMoE 1.68×, MoE-APEX 1.49×, and PDScope 1.38×. KTransformers exhibits a ratio of 2.86×, due to its CPU kernels optimized for Intel AVX-512, which do not generalize to AMD.

For Mixtral-8×7B, single-expert I/O cost far exceeds its CPU computation cost; KTransformers has a relative advantage due to its static placement and CPU-optimized kernels, while performance differences among other methods primarily stem from pipeline bubbles from I/O operations. For models with more but smaller experts, dynamic scheduling methods gain some advantage through load balancing, but PDScope further extends its lead via intelligent prefetching and minimal runtime overhead.

In summary, during decode, PDScope achieves up to 6.1× the throughput of Fate, 2.9× of KTransformers, 2.9× of Fiddler, 2.1× of HybriMoE, and 2.1× of MoE-APEX.

\begin{figure}[t!]
  \centering
  \begin{subfigure}[ht]{\linewidth}
    \includegraphics[width=\linewidth]{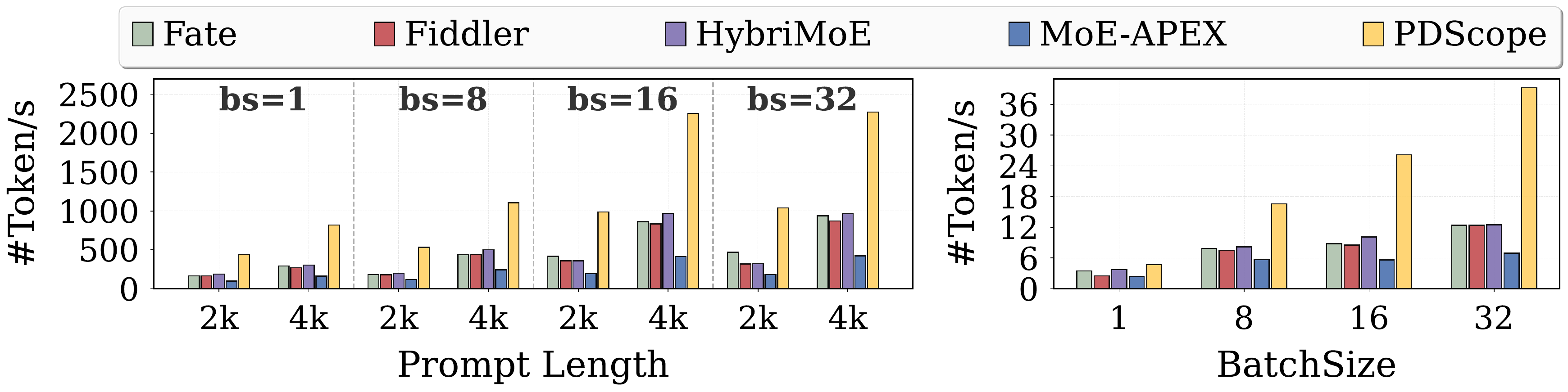}
  \end{subfigure}
\caption{Throughput on DeepSeek-V4-Flash under different batch sizes: (left) prefill; (right) decode.}
     \label{figurebs4} 
\end{figure}

\subsection{Scalability Analysis}
\label{subsec:scalability}

We evaluate PDScope's scalability on the A100 platform using DeepSeek-V4-Flash from two dimensions: model scale and batch scale, as shown in Fig.~\ref{figurebs4}.

\textbf{Model scalability.} KTransformers is excluded as it does not support this model on A100. MoE-APEX exhibits lower performance because it requires storing both full-precision and compressed expert copies on the CPU; enabling pinned memory would exceed the 512GB DRAM capacity, so this I/O optimization is disabled. PDScope achieves 2.82×, 3.06×, 2.69×, and 5.03× prefill throughput improvement over Fate, Fiddler, HybriMoE, and MoE-APEX, respectively, and 1.35×, 1.90×, 1.26×, and 1.97× decode throughput improvement in single-batch settings.

\textbf{Batch scalability.} In multi-batch, all systems enable chunked prefill (chunk size 256) to stay within the 24GB VRAM limit. At batch size 32, PDScope achieves 2.42×, 2.61×, 2.35×, and 5.37× prefill throughput improvement over Fate, Fiddler, HybriMoE, and MoE-APEX, respectively, and 3.16×, 3.17×, 3.14×, and 5.65× decode throughput improvement. Notably, increasing batch size from 16 to 32 yields only modest prefill throughput gains across all systems (e.g., PDScope: 5.2\% for 2048-token, 0.7\% for 4096-token), indicating that a batch size of 16 already saturates the hardware in this setting.

\begin{figure}[t]
  \centering
  \begin{subfigure}[ht]{\linewidth}
    \includegraphics[width=\linewidth]{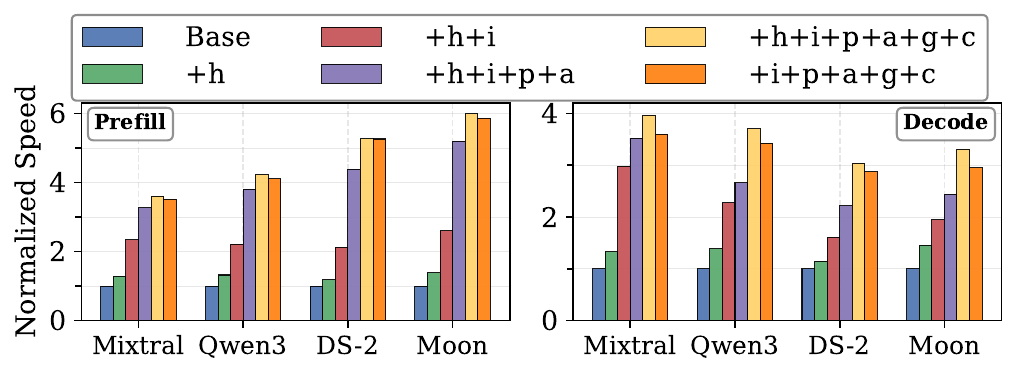}
  \end{subfigure}
\caption{Normalized speed compared to baseline: \texttt{+h} (hot expert table), \texttt{+i} (AsyncIO), \texttt{+p} (PhasePer), \texttt{+a} (AdaptSched), \texttt{+g} (static graph), \texttt{+c} (cache prefetching)}
     \label{figure13} 
\end{figure}

\subsection{Ablation Study}
\label{subsec:ablation}

To quantify the contribution of each component, we start from Fiddler as the baseline on the A100 platform and incrementally enable PDScope's optimizations. Results are shown in Fig.~\ref{figure13}. All reported improvements are incremental gains computed as the ratio of throughput with the component enabled to throughput without it.

\textbf{Hot expert table (+h).} This optimization trades spare GPU memory for reduced on-demand I/O overhead by preloading the most frequently accessed hot experts, delivering 1.16×–1.39× and 1.15×–1.46× throughput improvement in prefill and decode, respectively.
\textbf{AsyncIO (+i).} This component reduces the effective I/O overhead of expert data transfer, delivering 1.67×–1.88× in prefill and 1.34×–2.24× in decode.
\textbf{PhasePer \& AdaptSched (+p+a).} These core components are evaluated together, as PhasePer's predictions are the direct input to AdaptSched's scheduling decisions; separating them would break this causal chain and underestimate their combined effect. Their synergy delivers 1.39×–2.06× throughput in prefill by exploiting predictable load imbalance. Decode-phase gains range from 1.17× to 1.38× across models.
\textbf{Static graph \& cache prefetching (+g+c).} These system-level optimizations deliver 1.10×–1.20× in prefill and 1.13×–1.39× in decode, with more pronounced gains in decode where fine-grained runtime overheads dominate.

\textbf{Robustness under memory constraints.} Disabling the hot expert table results in only negligible prefill throughput loss (avg. 1.8\%) and modest decode loss (avg. 8.2\%), reflecting functional overlap with other components. In this configuration, expert weights are streamed directly through the I/O buffers described in \S\ref{secio}, overwriting previous content without explicit eviction.

\subsection{Predictor Accuracy}
\label{preacc}

We evaluate PhasePer's generalization on three datasets (ShareGPT, DPO, LLaVA) using 4096-input/512-output samples, after offline training on 16\% of ShareGPT. Prediction accuracy is measured as the hit rate against the experts selected by the gating network.

\begin{figure}[t!]
  \centering
  \begin{subfigure}[ht]{\linewidth}
    \includegraphics[width=\linewidth]{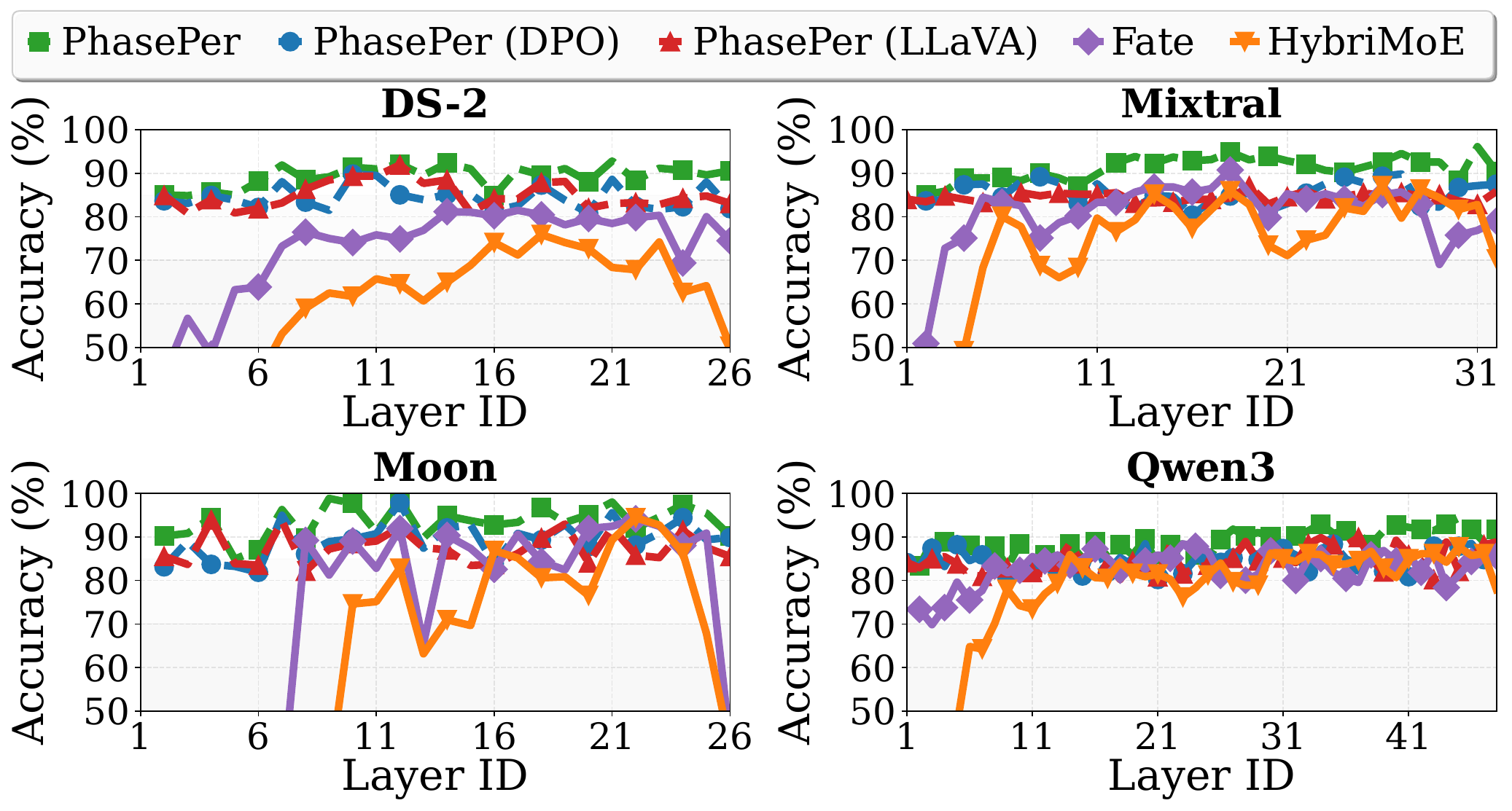}

  \end{subfigure}
\caption{Prefill-phase prediction accuracy of PhasePer across different models and datasets.}
     \label{figure14} 
\end{figure}

Fig.~\ref{figure14} shows the results. PhasePer's accuracy on out-of-domain datasets (DPO, LLaVA) drops only modestly (avg. 4.8–5.3\%) compared to ShareGPT, and never falls below 80\% even in low-similarity input/output groups—demonstrating strong generalization to domain shifts. Compared to Fate, PhasePer improves accuracy by 15\% to 68.4\%. Decode-phase results follow the same trend, with average accuracy 2.3\% higher than in prefill due to stronger inter-layer similarity; full decode data are in the \textit{Appendix} \S\ref{a2}.

\begin{table}[t]
\centering
\caption{Performance Evaluation of PhasePer.}
\label{tab:model_performance}
\resizebox{\linewidth}{!}{%
\begin{tabular}{l *{4}{>{\centering\arraybackslash}p{2cm}}}
\toprule
 & \textbf{DeepSeek} & \textbf{Mixtral} & \textbf{Qwen} & \textbf{Moonlight} \\
\midrule
\textbf{ShareGPT Top-6} & 96\% & 100\% & 97\% & 99\% \\
\textbf{DPO Top-6} & 92\% & 99\% & 94\% & 95\% \\
\textbf{LLaVA Top-6} & 90\% & 99\% & 92\% & 92\% \\
\hline
\textbf{Memory} & 1.6-4.0 MB & 1.2-3.8 MB & 1.7-4.2 MB & 1.6-4.0 MB \\
\textbf{Time} & 0.48 ms & 0.15 ms & 0.12 ms & 0.48 ms \\
\bottomrule
\end{tabular}%
}
\label{tb3}
\end{table}

We further examine PhasePer's impact on end-to-end performance. Table~\ref{tb3} reports the Top-6 prefetch accuracy for experts ranked by token count, where a prediction is correct if the predicted expert falls within a ±2 position range of the true ranking (e.g., the predicted $6^{th}$ expert is within the true $4^{th}$ to $8^{th}$). This tolerance-based metric is chosen because AdaptSched's scheduling decisions rely on the relative ordering of expert "hotness" rather than exact rankings; small positional deviations still guarantee the expected prefetching utility. PhasePer's Top-6 hit rate remains above 90\%, fully meeting AdaptSched's precision requirement.

In terms of resource usage, PhasePer deploys one micro-predictor per layer, with parameter counts as low as 1.2 MB; middle-layer predictors use slightly deeper architectures, up to 4.2 MB. The forward-pass latency ranges from 0.12 to 0.48 ms; this latency is hidden within GPU idle time during expert transfers, introducing no additional pipeline bubbles.

\begin{figure}[t]
  \centering
  \begin{subfigure}[ht]{\linewidth}
    \includegraphics[width=\linewidth]{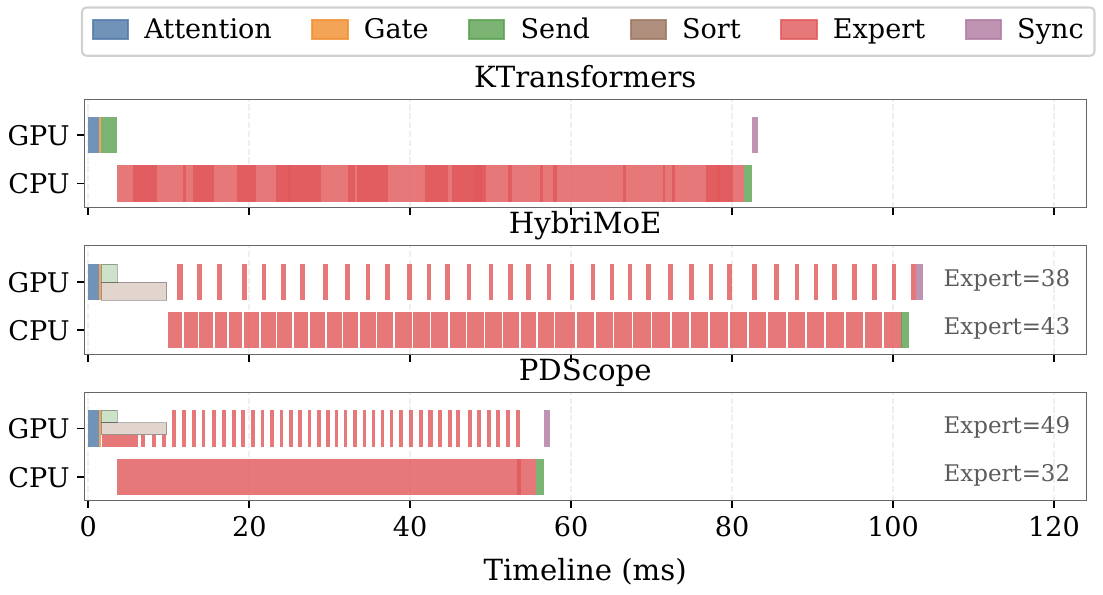}
    \caption{Prefill phase}
    \label{figure15a}
  \end{subfigure}
  \hfill
  \begin{subfigure}[ht]{\linewidth}
    \includegraphics[width=\linewidth]{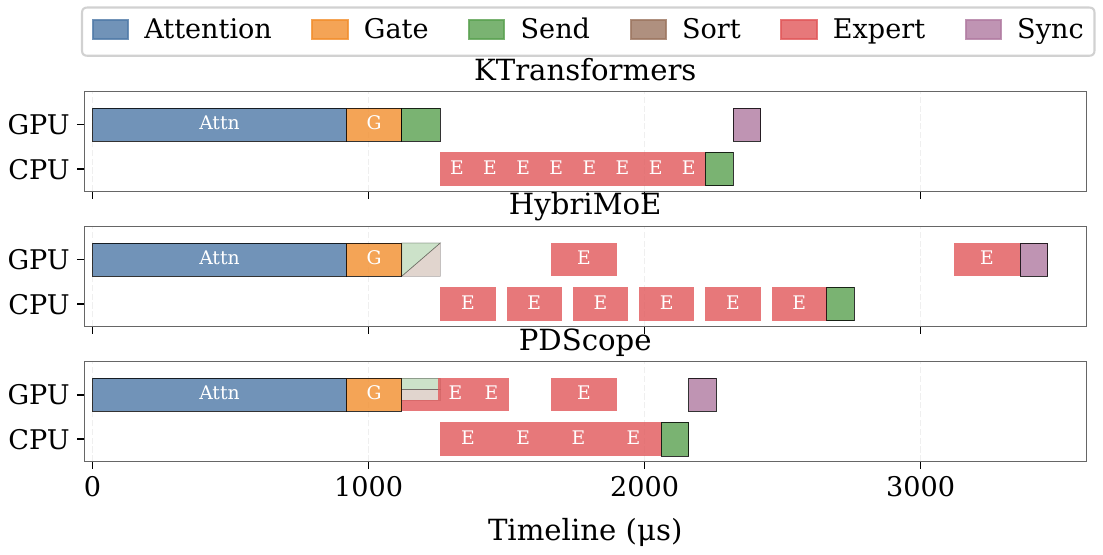}
    \caption{Decode phase}
    \label{figure15b}
  \end{subfigure}
  \caption{Runtime breakdown on Qwen3-30B-A3B (A100).}
  \label{figure15}
\end{figure}

\subsection{Runtime Trace Analysis}
\label{subsec:micro-trace}

To gain a microscopic understanding of the performance differences, we capture runtime execution traces of a single layer of Qwen3-30B-A3B on A100 with 1024 input tokens and 64 output tokens (Fig.~\ref{figure15}). Hot expert preloading is disabled to avoid confounding with runtime prefetching.

In prefill, KTransformers executes all expert computation on the CPU, with performance limited by CPU compute capability. HybriMoE, despite its dynamic load-balancing capability to reduce GPU idleness, does not achieve superior performance. In contrast, PDScope's trace shows higher GPU utilization. Computation gaps are smaller because the static graph eliminates the dynamic scheduler's polling overhead and AsyncIO improves I/O transfer performance. A higher proportion of experts are also prefetched per layer, reflecting PhasePer's accurate predictions and AdaptSched's cross-layer coordination. The "Sort"—which encompasses both scheduling decisions and graph construction—takes 8.1 ms and overlaps with 12 GPU expert computations and 4 CPU expert computations, confirming that these operations can be hidden within the pipeline without adding extra delay.

In decode, KTransformers runs efficiently on CPU with optimized kernels and no scheduling overhead. HybriMoE's second expert transfer does not arrive at the GPU at the expected time, failing to achieve ideal load balancing. PDScope improves system performance through three mechanisms: (1) PhasePer ensures prefetched experts are used, reducing bandwidth waste; (2) AdaptSched's decode-phase scheduling is more precise, avoiding performance loss from excessive I/O; and (3) system optimizations such as kernel launch batching and cache prefetching ensure data readiness before computation begins, reducing computation gaps.

These traces validate PDScope's core claims: phase-adaptive scheduling eliminates greedy underfetching in prefill and aggressive I/O in decode; system-level optimizations translate decisions into performance without runtime overhead.


\section{Discussion}
This paper targets low-batch inference as the canonical model for studying phase heterogeneity. PDScope does not support the prefill/decode overlap used in distributed inference: on memory-constrained CPU-GPU nodes, prefill bottlenecks shift to CPU and I/O (\S\ref{sec22})—overlapping decode would only intensify contention. PDScope supports multi-batch phase-consolidated batching (\S\ref{subsec:scalability}): AdaptSched reuses the same prefill logic, while decode operates in Mode d, where operations are planned without resource contention.

For platform diversity, the offline cost table calibrates operation latencies, enabling AdaptSched to adapt to different hardware. The full offline phase takes under 30 minutes per model on A100. AdaptSched is resilient to prediction errors: even when PhasePer accuracy drops to 45–70\%, prefill throughput decreases by only 6.4–11.3\%. Step 2's layer-local re-evaluation prevents cascading errors by falling back to on-demand loading for mispredicted experts. Detailed analysis is in the \textit{Appendix} \S\ref{{erro}}. 

The hot expert table is derived from the offline phase, a static cache with limited capacity under tight memory. Disabling it in §\ref{subsec:ablation} has limited impact on performance.


\section{Conclusion}

This paper identifies phase heterogeneity—the fundamentally different compute and I/O demands of prefill and decode—as the root cause of inefficiency in MoE inference on memory-constrained CPU-GPU nodes. Existing phase-agnostic systems, regardless of their scheduling strategy, inevitably compromise one phase for the other. PDScope demonstrates that explicitly separating the two phases across prediction, scheduling, and execution layers turns this heterogeneity from a liability into a schedulable resource.
Our evaluation shows that PDScope achieves up to 10.3× prefill and 6.1× decode throughput over state-of-the-art baselines. 


\begin{acks}
This work was supported in part by the New Generation Artificial Intelligence-National Science and Technology Major Project (Grant No. 2025ZD0122202), the National Natural Science Foundation of China (Grant Nos. 62421002 and U24B20151), and the Science and Technology Innovation Program of Hunan Province (Grant No. 2025RC3118). Dezun Dong, Shaonan Ma, Yongwei Wu, and Xiangke Liao hold joint appointments at Qiyuan Lab. Dezun Dong and Zhaoning Zhang are the corresponding authors.
\end{acks}

\bibliographystyle{ACM-Reference-Format}

\bibliography{bib2}
\section{appendix}

\begin{figure}[ht]
  \centering
  \begin{subfigure}[ht]{\linewidth}
    \includegraphics[width=\linewidth]{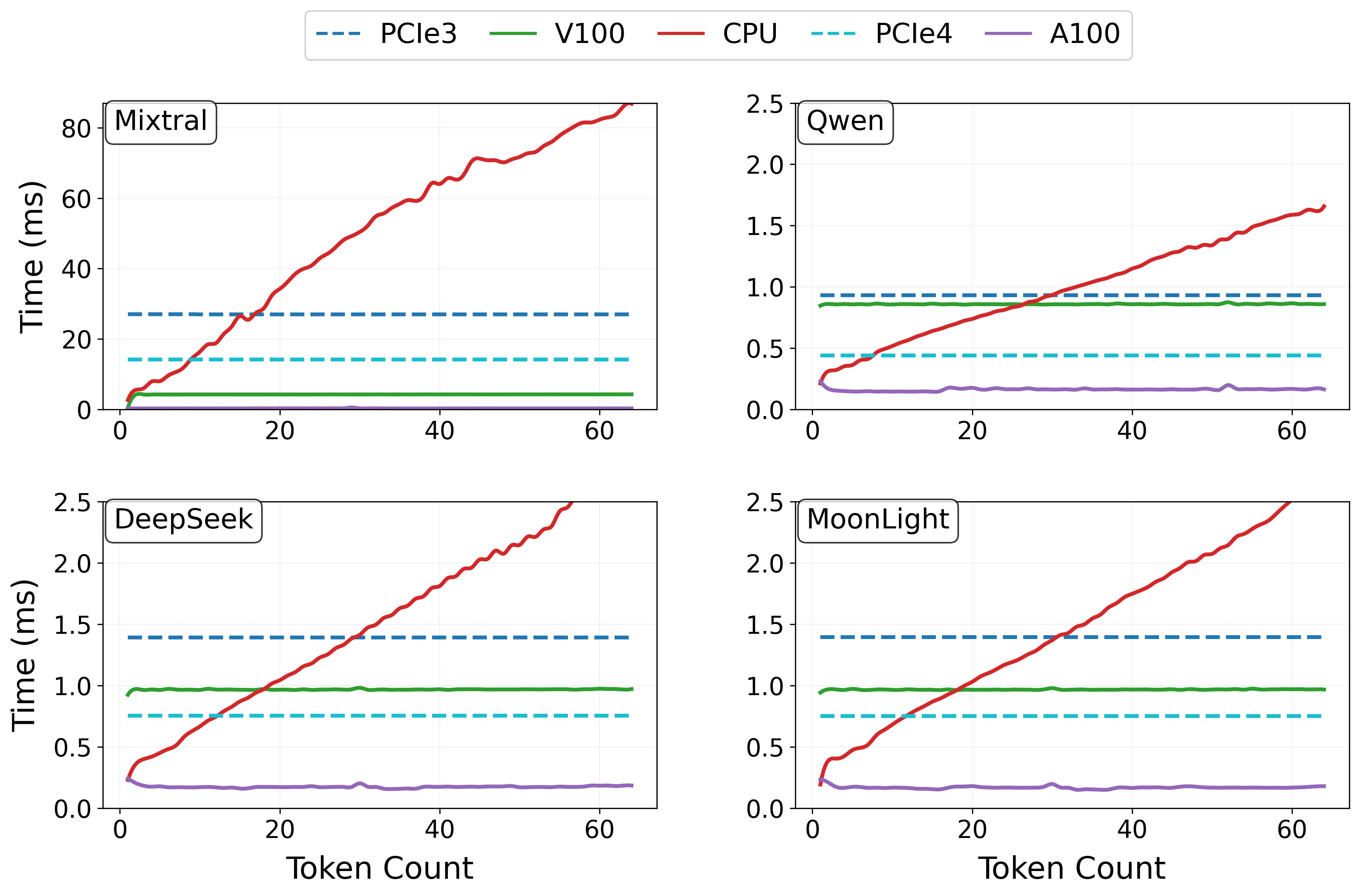}
  \end{subfigure}
  \caption{Characterization of expert computation and transfer costs across models and heterogeneous devices.}
     \label{figure18} 
\end{figure}

\subsection{Overhead Characterization}
\label{A1}

Fig.~\ref{figure18} reveals three key observations. First, for both V100 and A100 GPUs, single-expert GPU execution time is significantly lower than PCIe transfer time, indicating that GPU-only solutions are I/O-bound rather than compute-bound. Second, for small-expert architectures (e.g., Qwen), CPU computation delay is lower than GPU startup delay when tokens per expert is below 16; for large-expert models like Mixtral, CPU delay consistently exceeds GPU, determining the break-even point for CPU offloading. Third, as token count increases, GPU's parallel advantage becomes prominent while CPU overhead grows linearly, suggesting GPU should dominate large-volume computation.

\begin{figure}[ht]
  \centering
  \begin{subfigure}[ht]{0.49\linewidth}
    \includegraphics[width=\linewidth]{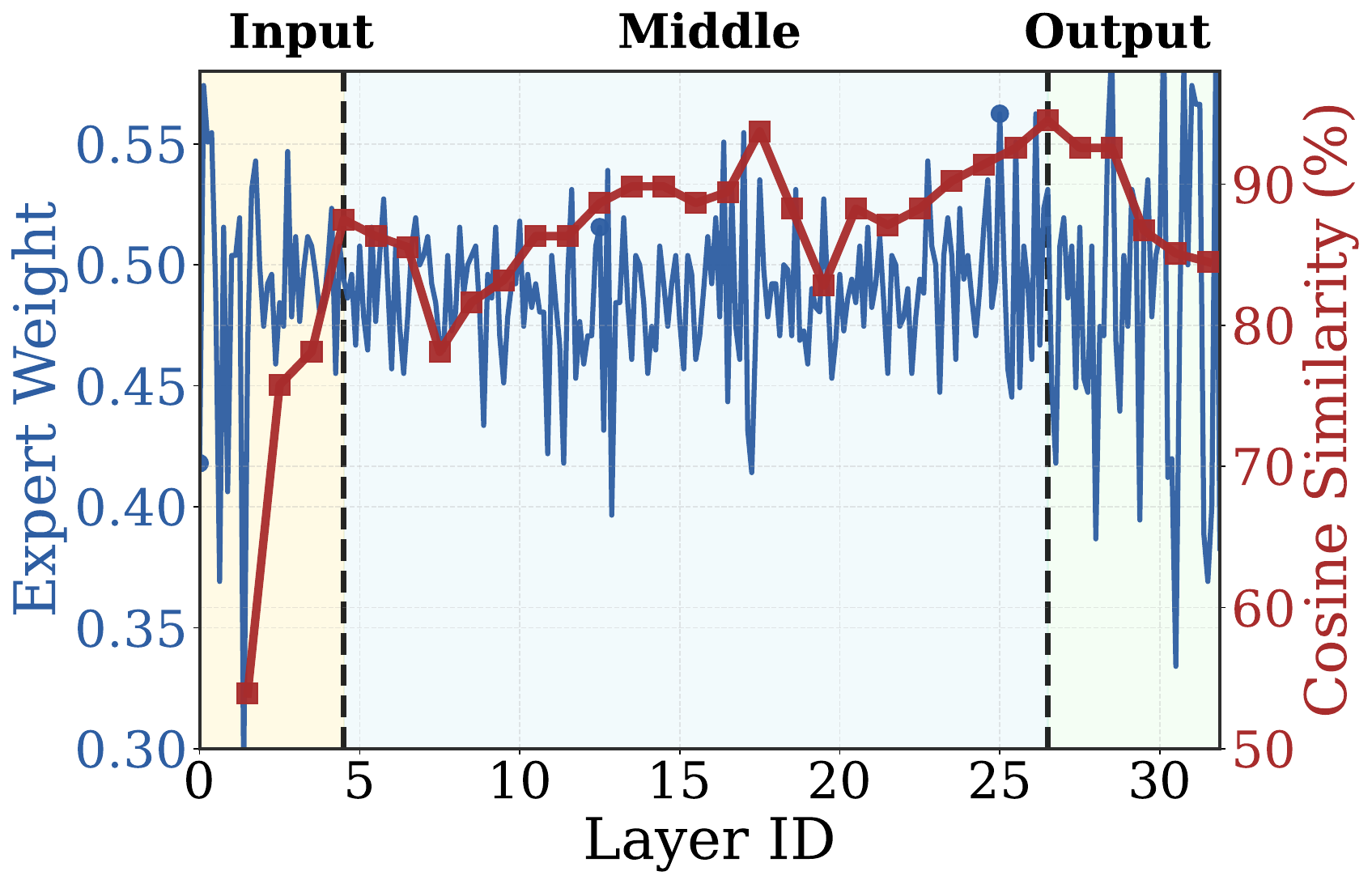}
  \caption{Mixtral}    
  \end{subfigure}
  \begin{subfigure}[ht]{0.49\linewidth}
    \includegraphics[width=\linewidth]{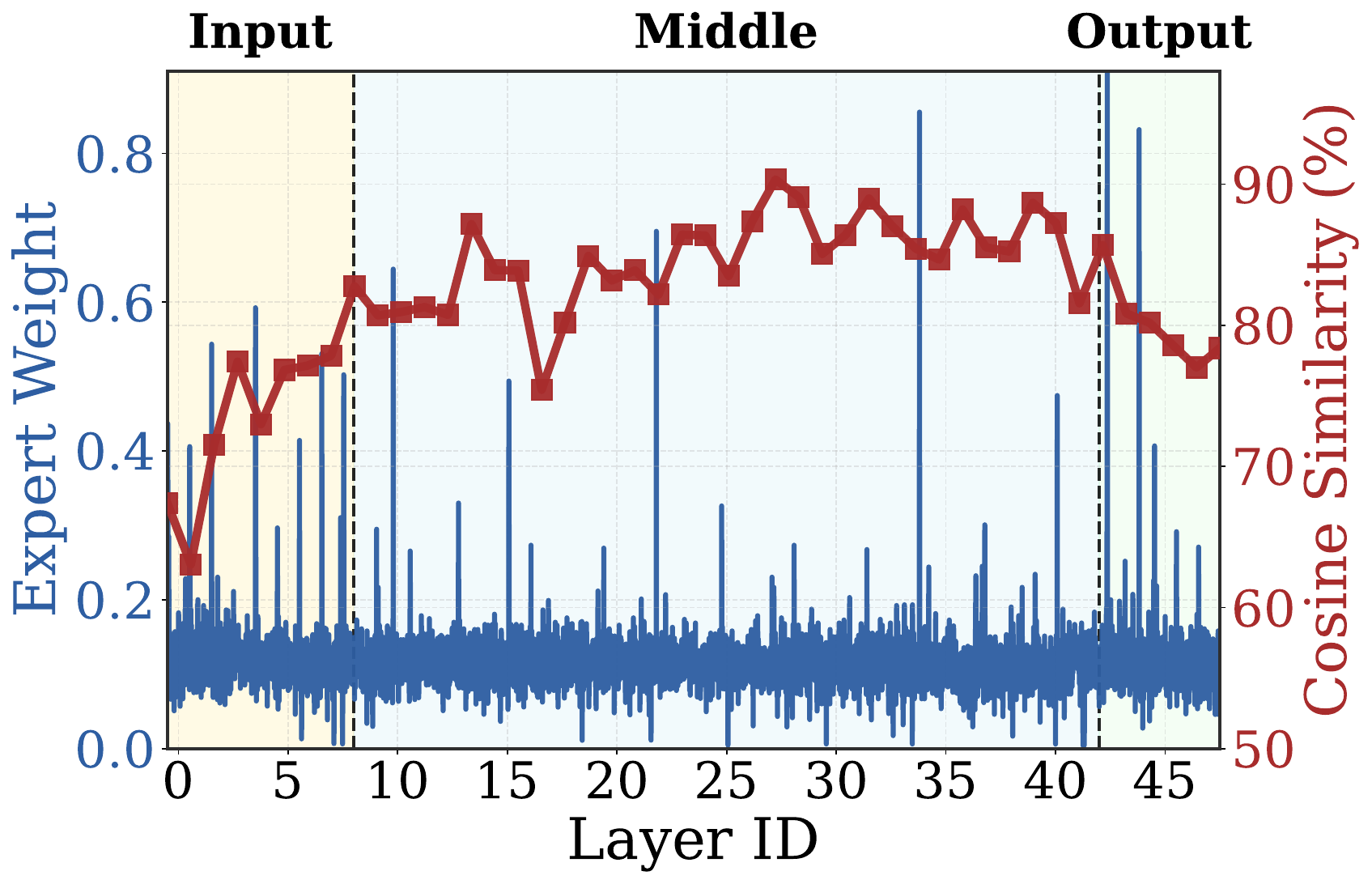}
  \caption{Qwen3}    
  \end{subfigure}
    \begin{subfigure}[ht]{0.49\linewidth}
    \includegraphics[width=\linewidth]{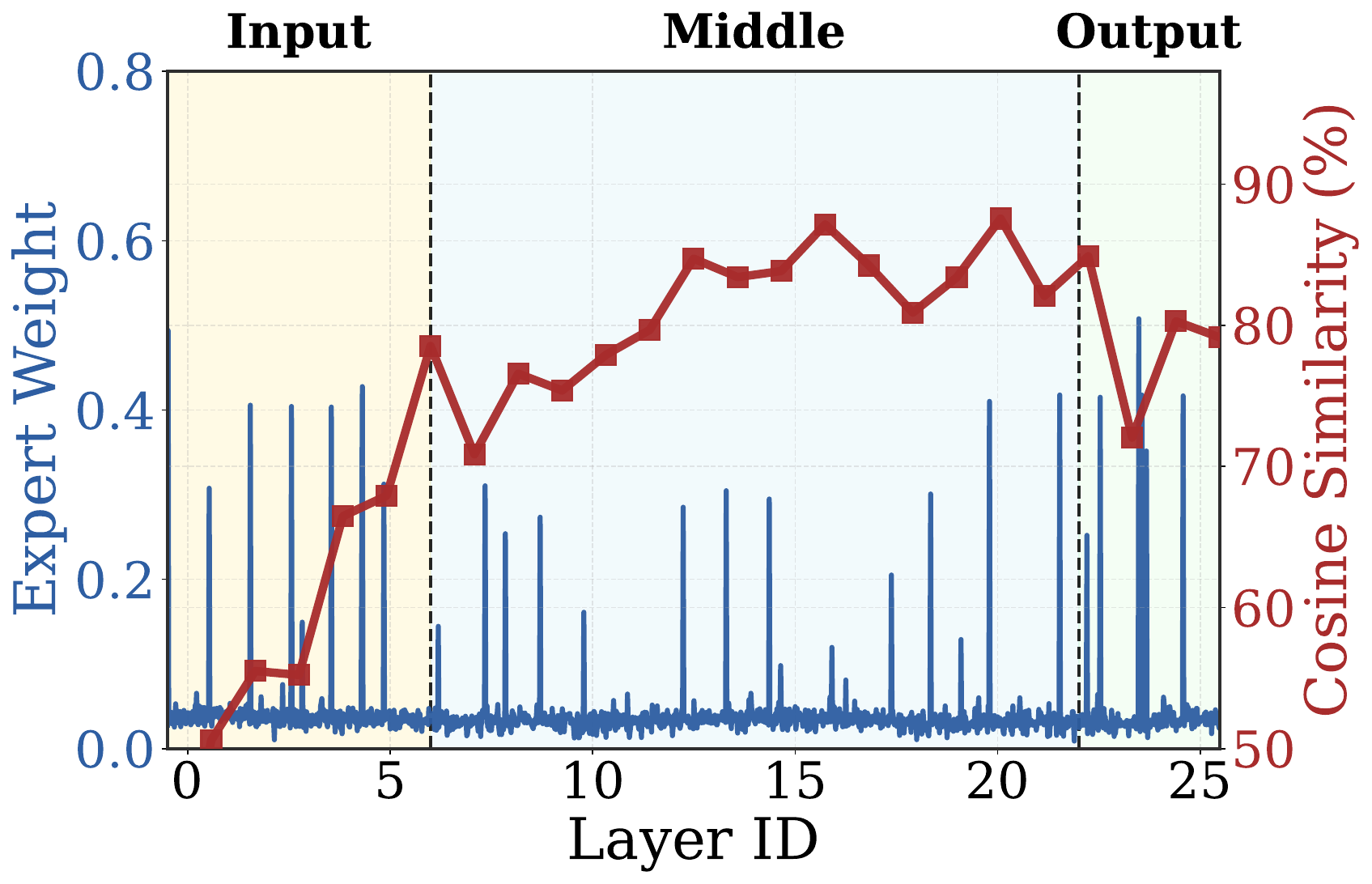}
  \caption{DeepSeek}    
  \end{subfigure}
    \begin{subfigure}[ht]{0.49\linewidth}
    \includegraphics[width=\linewidth]{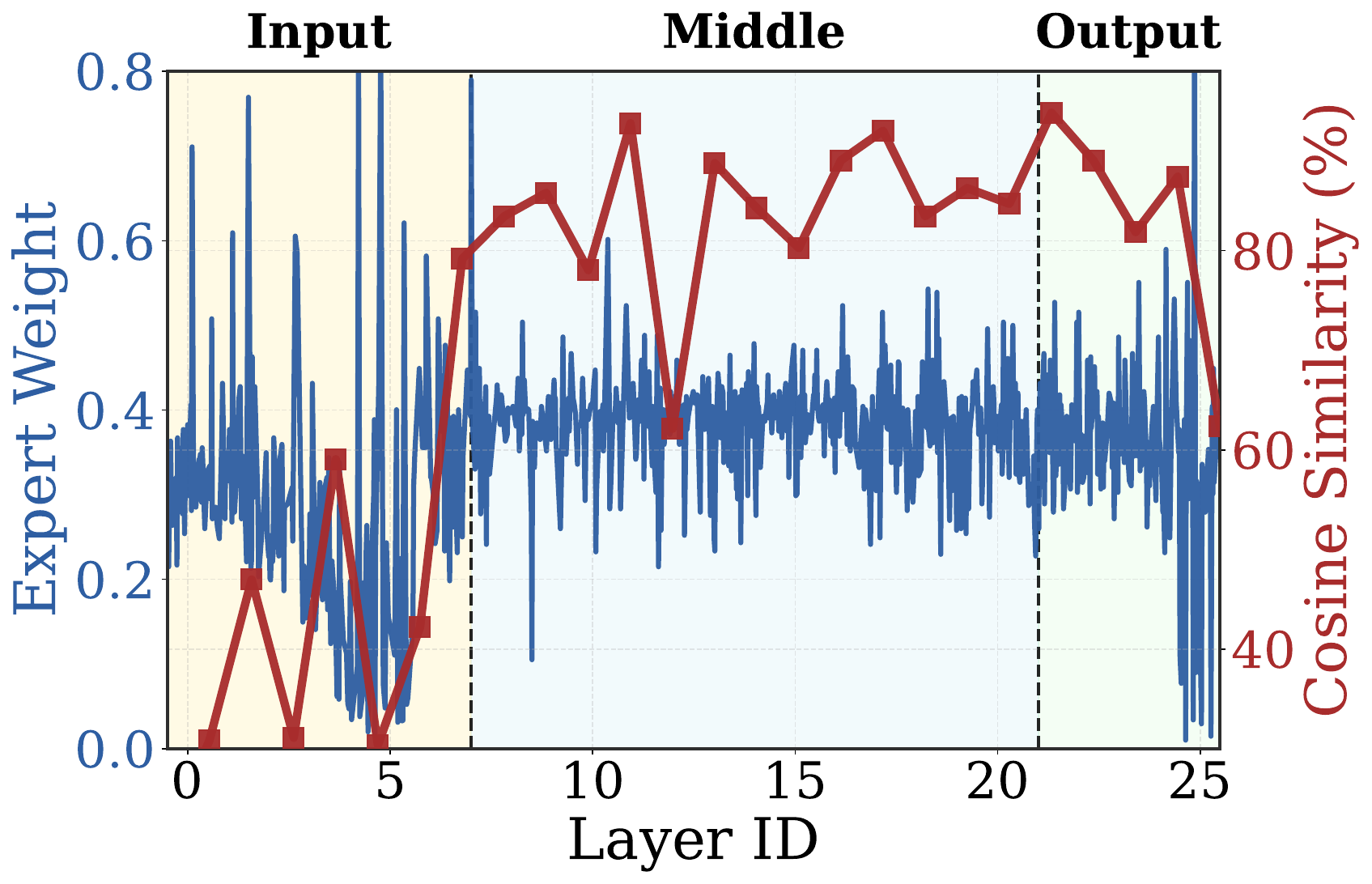}
  \caption{Moonlight}    
  \end{subfigure}
  \caption{Group-wise patterns in routing weights and inter-layer similarity in MoE models.}
     \label{figure21} 
\end{figure}
\subsection{Layer-group Effects of Different Models}
\label{a2}
\begin{figure}[t]
  \centering
  \begin{subfigure}[ht]{\linewidth}
    \includegraphics[width=\linewidth]{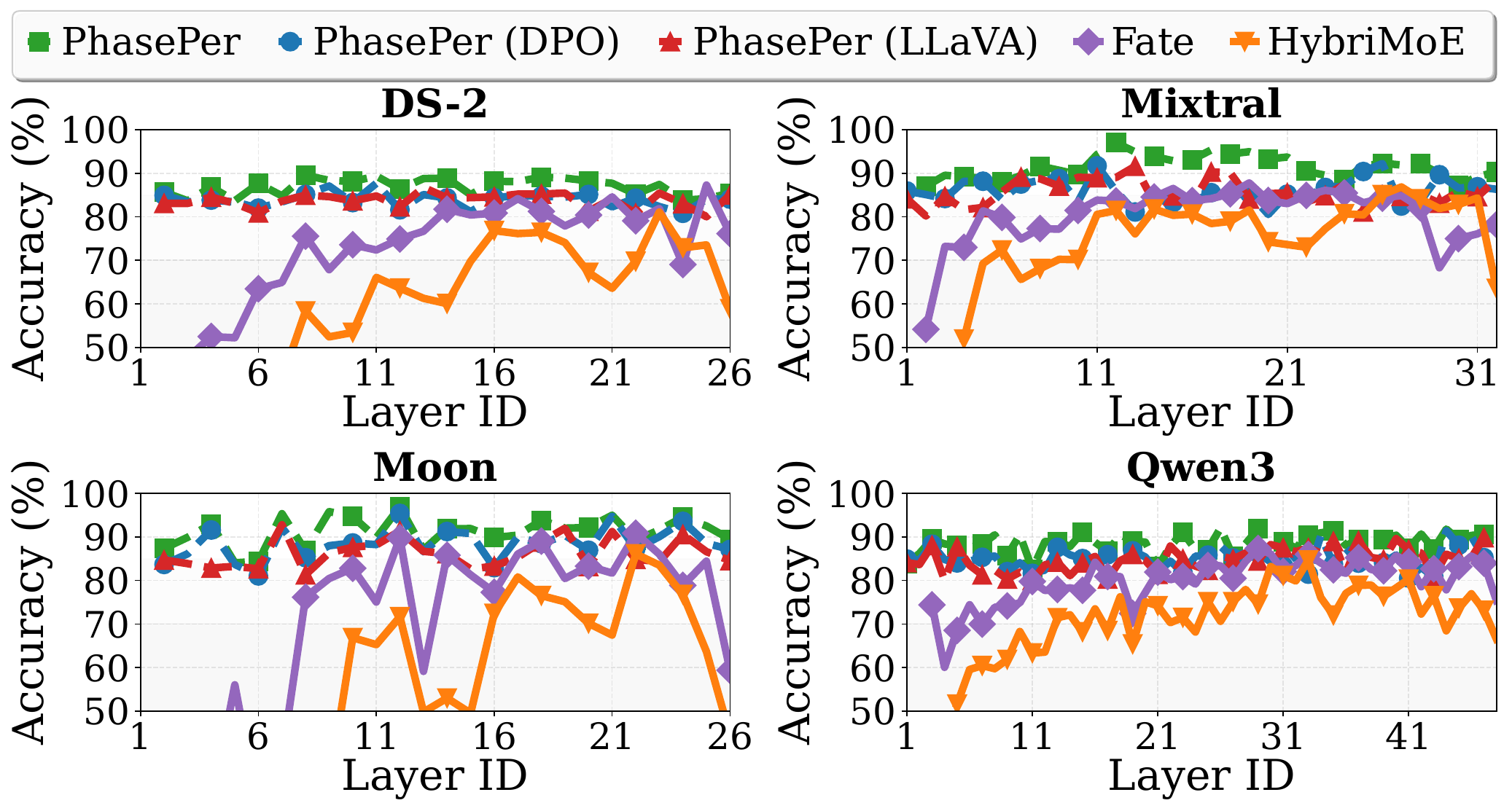}
  \end{subfigure}
  \caption{Accuracy of PhasePer in the decode phase.}
     \label{figure24b} 
\end{figure}

Fig.~\ref{figure21} shows that all models exhibit significant fluctuation in routing weights within input/output layer groups, indicating strong inter-layer expert correlations in these regions. Within middle layer groups, inter-layer cosine similarity remains consistently high, enabling stable prediction accuracy based on attention-output hidden states. The boundaries between layer groups can be automatically identified by inflection points in cosine similarity: a monotonic increase from low to high similarity during input-to-middle transition, and a corresponding decrease during middle-to-output transition. This observation provides the foundation for PhasePer's adaptive layer-grouping. Decode-phase prediction accuracy for PhasePer is provided in Fig.~\ref{figure24b}.

\subsection{Robustness to Prediction Errors}
\label{erro}

Although PhasePer demonstrates strong generalization capability in cross-dataset tests (\S\ref{preacc}), we further evaluate the potential impact of predictor accuracy degradation on AdaptSched's performance under extreme conditions. For this purpose, we replaced PhasePer with the prediction strategy used by Fate, which directly predicts the next layer's expert activation based on the previous layer's hidden state passed through the gating module.

Tests on an A100 GPU with an input length of 4096 and output length of 64 showed that prefill throughput for the input/output layer groups decreased by an average of 6.4\%–11.3\%, while the middle layer group decreased by only 0.9\%–2.4\%. Combined with the ablation study results (\S\ref{subsec:ablation}), these findings demonstrate that even with lower prediction accuracy, AdaptSched's scheduling performance remains significantly superior to layer-by-layer on-demand scheduling strategies without prediction. This robustness under prediction inaccuracy stems from the interaction between AdaptSched's design and the statistical properties of expert activation in sparse MoE models.

\textbf{Statistical basis of robustness.} The fundamental logic can be understood through the relationship between the random baseline hit rate and the ranking advantage conferred by prediction. We use Mixtral-8×7B as an example. Under the (unrealistic) assumption of random expert selection—ignoring all routing preferences—the probability of any given expert being selected for a token is $2/8 = 25\%$. This is the random baseline hit rate.

\textbf{Why ranking matters more than perfect accuracy.} For a predictor to deliver systematic prefetching benefits, it does not need perfect accuracy. Instead, it only needs to achieve a hit rate for the true "hot experts" (those that ultimately process the most tokens) that is consistently higher than the 25\% random baseline. Within a large batch, expert activation can be viewed as multiple independent Bernoulli trials. The value of the predictor lies in generating a ranking of expert "hotness" that is more accurate than random. Even with moderate absolute accuracy, if the predictor's hit rate for the true hot experts exceeds the baseline, those experts are more likely to appear higher in the predicted ranking. Consequently, prefetching I/O operations are statistically more likely to perform "useful work," improving overall throughput.

\textbf{System-level safeguard.} If a hot expert is not prefetched, AdaptSched's layer-local re-evaluation mechanism defaults to on-demand loading for that expert, limiting the latency penalty of any single prediction error. This explains AdaptSched's robustness even with relatively simple predictors.

\textbf{Implications for different models.} For models with small expert counts and low I/O costs, the direct overhead of prediction errors is low, and this mechanism effectively absorbs inaccuracies. For models like Mixtral-8×7B, where each layer has 8 experts under Top-2 activation, most experts have a considerable probability of being used during long-prompt processing. Thus, even prefetching a non-"hottest" expert can often yield partial performance benefits.

\subsection{Fine-Tuning Dynamics and Error Propagation}

We further analyze two practical aspects that complement the above robustness discussion.

\textbf{Error propagation prevention.} Even in the rare case of consecutive mispredictions, AdaptSched's layer-local re-evaluation (Step~2 in Algorithm~\ref{alg:AdaptSched}) acts as a safety net. When PhasePer incorrectly predicts the top-$k$ experts for layer $l+1$, the scheduler may prefetch experts that are never used. However, Step 2 re-evaluates the on-demand loading boundary for the current layer $l$ based on actual token counts, ignoring prefetched but unused experts. For the affected layer $l+1$, the missing experts are fetched on-demand, incurring at most one extra I/O round trip per mispredicted layer. This prevents cascading errors across multiple layers.

\textbf{Online fine-tuning trigger frequency and latency hiding.} The error threshold over a 16-token sliding window is conservatively low. On all evaluated datasets (ShareGPT, DPO, LLaVA), the average interval between fine-tuning triggers exceeds 500 inference steps for prefill and over 1000 steps for decode, because prediction errors are rarely consecutive. The fine-tuning overhead (2.1 ms per trigger) is entirely overlapped with GPU idle periods that naturally occur during expert I/O transfers. In the decode phase, fine-tuning is scheduled during the attention computation of the next token, again achieving full overlap. Consequently, fine-tuning adds zero observable delay to the critical path.

\begin{figure}[ht]
  \centering
  \begin{subfigure}[ht]{\linewidth}
    \includegraphics[width=\linewidth]{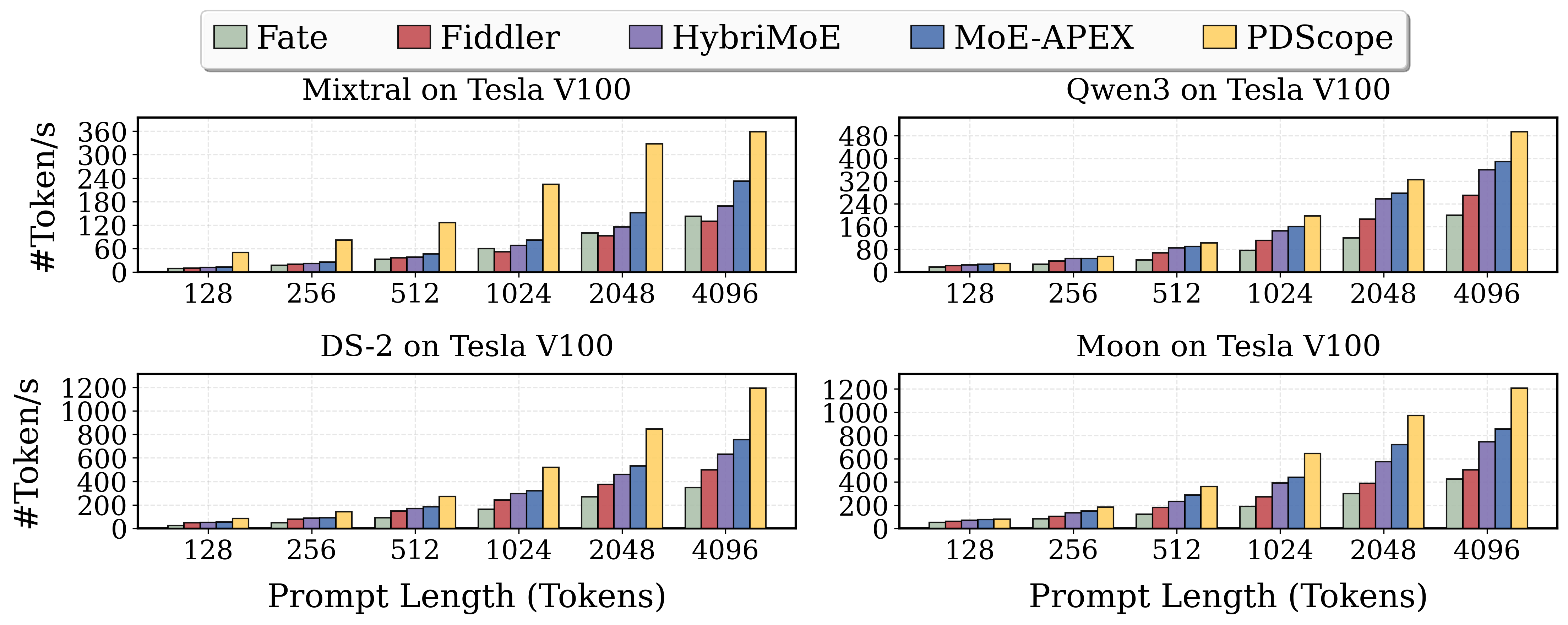}
  \end{subfigure}
  \caption{Throughput comparison of PDScope and baseline systems in the prefill phase.}
     \label{figure19} 
\end{figure}

\subsection{Throughput on V100}
\label{A3}
\subsubsection{Prefill Throughput}
\begin{figure}[ht]
  \centering
  \begin{subfigure}[ht]{\linewidth}
    \includegraphics[width=\linewidth]{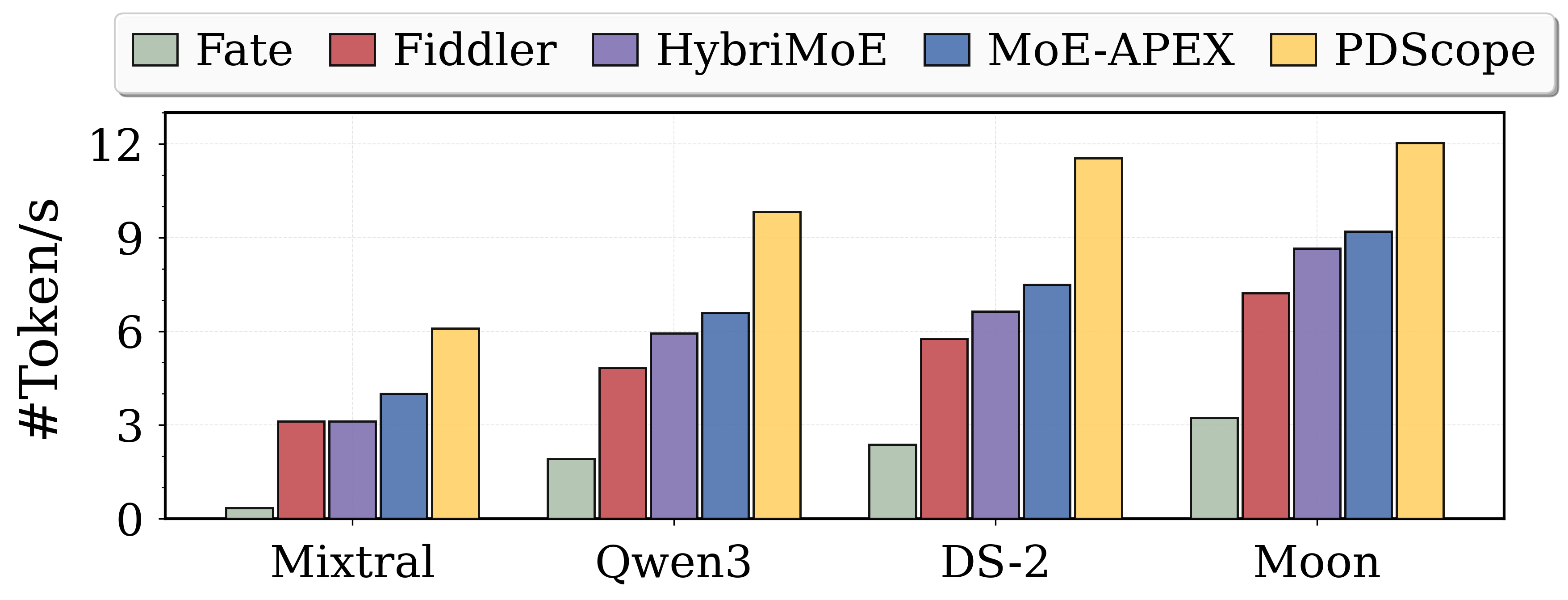}
  \end{subfigure}
  \caption{Throughput comparison of PDScope and baseline systems in the decode phase.}
     \label{figure20} 
\end{figure}

We evaluate on an NVIDIA V100 GPU (PCIe 3.0 x16) with an Intel Xeon Gold 6230R CPU. KTransformers is excluded as it does not support pre-Ampere GPUs. Results in Fig.~\ref{figure19} and Fig.~\ref{figure20} show that relative performance ranking shifts only modestly. All systems show significant performance drop on Qwen3 due to: (1) higher PCIe 3.0 transfer overhead, and (2) higher kernel launch overhead on V100 vs. A100. Consequently, more computation shifts to CPU, limiting overall efficiency.

\subsubsection{Decode Throughput}
PDScope maintains its performance advantage on V100, consistent with the analysis in §\ref{sec22}: decode-phase compute intensity is inherently low, and the core scheduling requirements are avoiding I/O overhead and scheduling-induced latency. Since PDScope's key optimizations—PhasePer, AdaptSched, and system-level optimizations—are not tied to specific hardware capabilities, reduced GPU and PCIe performance has limited impact on its relative gains.

\begin{figure}[ht]
  \centering
  \begin{subfigure}[ht]{\linewidth}
    \includegraphics[width=\linewidth]{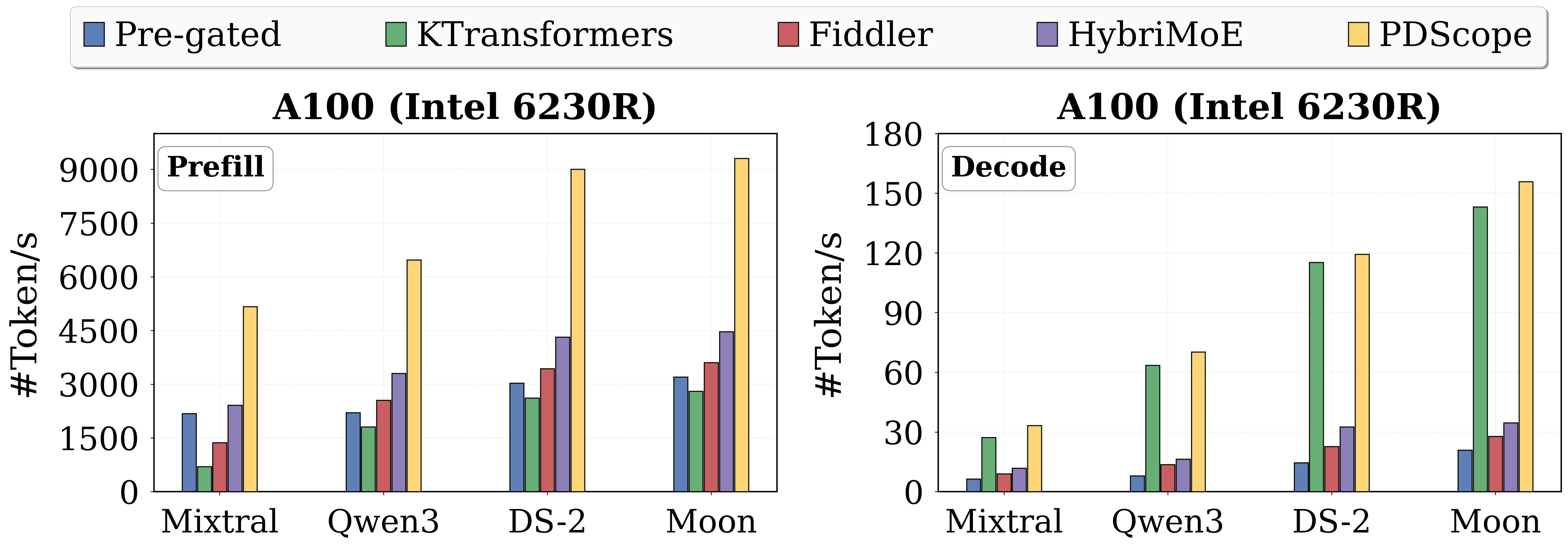}
  \end{subfigure}
  \caption{Throughput comparison of PDScope and baseline systems with a batch size of 32. }
     \label{figure202} 
\end{figure}



\subsection{Multi-Batch Inference Performance}
\label{A33}

We complement the scalability analysis in §\ref{subsec:scalability} with additional observations on A100 at batch size 32 (1024 input, 128 output tokens), using the full set of evaluated models. Results in Fig.~\ref{figure202} confirm that the seesaw effect persists under concurrency: KTransformers excels in decode while Fate excels in prefill, with reversed trends—consistent with the phase-agnostic limitation discussed in §\ref{sec:introduction}. PDScope's decode-phase scheduler directly employs Mode d (cross-layer focus) from Algorithm~\ref{alg:AdaptSched}, while prefill-phase gains align with the long-context single-batch results in Fig.~\ref{figure11}. In decode, KTransformers benefits from AVX-optimized kernels on Intel platforms, yet PDScope further improves throughput via systematic overhead reduction and adaptive load balancing, validating that phase-aware scheduling remains effective even when inter-request parallelism is available.





\subsection{The Lookahead Depth of AdaptSched}
\label{look}

A key question for AdaptSched's cross-layer scheduling is whether looking ahead only one layer (i.e., layers $l$ and $l+1$) sufficiently captures the available optimization potential, or if a longer horizon (e.g., layers $l$, $l+1$, and $l+2$) is necessary. To investigate this, we compared a two-layer lookahead strategy against the default one-layer lookahead.

The two-layer strategy faces a practical limitation. If layers $l$ and $l+1$ are both dominated by hot experts under cross-layer visibility, while layer $l+2$ is dominated by cold experts, the CPU has insufficient work to remain busy, and PCIe has little idle time to prefetch experts for layer $l+2$. In this case, the two-layer strategy must still degrade to one-layer lookahead to ensure optimal expert placement for layers $l$ and $l+1$, rather than leaving the CPU idle across both layers. In practice, the two-layer strategy yields tangible benefits only when layers $l$ and $l+1$ are both dominated by cold experts and layer $l+2$ by hot experts—a scenario where prefetching for layer $l+2$ can be effectively overlapped with CPU-bound computation of the preceding two layers. Furthermore, two-layer lookahead requires maintaining prefetch caches for experts across three layers simultaneously, imposing additional memory pressure.

Experiments on an A100 GPU with 4096-input and 64-output tokens show that the throughput difference between one- and two-layer lookahead is only 1.1\%–2.3\% for prefill and 0.8\%–5.4\% for decode across the four models. These results indicate that in long-context scenarios, one-layer lookahead effectively resolves the core scheduling questions—"whether to prefetch" and "how many to prefetch"—and delivers near-optimal performance.

\end{document}